\documentclass[aps,prl,twocolumn,a4paper,10pt,notitlepage,footinbib,superscriptaddress,showpacs,floatfix]{revtex4-1}%
\usepackage{graphicx,color}
\usepackage{amsmath,amssymb,latexsym}
\usepackage{mathrsfs}
\begin{document}

\title{Simplifying the Topology of Knotted Ring Polymers \\ by Entropic Competition of Diffusing Slip-Links}

\title{Simplifying Topological Entanglements by Entropic Competition of Slip-Links}

\author{Andrea Bonato}
\address{University of Edinburgh, SUPA, School of Physics and Astronomy, Peter Guthrie Road, EH9 3FD, Edinburgh, UK }

\author{Davide Marenduzzo}
\address{University of Edinburgh, SUPA, School of Physics and Astronomy, Peter Guthrie Road, EH9 3FD, Edinburgh, UK }

\author{Davide Michieletto}
\affiliation{University of Edinburgh, SUPA, School of Physics and Astronomy, Peter Guthrie Road, EH9 3FD, Edinburgh, UK }
\affiliation{MRC Human Genetics Unit, Institute of Genetics and Molecular Medicine, University of Edinburgh, Edinburgh EH4 2XU, UK}


\begin{abstract}
\textbf{Topological entanglements are abundant, and often detrimental, in polymeric systems in biology and materials science. Here we theoretically investigate the topological simplification of knots by diffusing slip-links (SLs), which may represent biological or synthetic molecules, such as proteins on the genome or cyclodextrines in slide-ring gels. We find that SLs entropically compete with knots and can localise them, greatly facilitating their downstream simplification by transient strand-crossing. We further show that the efficiency of knot localisation strongly depends on the topology of the SL network and, informed by our findings, discuss potential strategies to control the topology of biological and synthetic materials.}
\end{abstract}
\maketitle

Knots and topological entanglements are often found in physical and biological systems~\cite{history_science_knots,VanRijssel1990,Chen1995,Wasserman1985a}.
Their uncontrolled formation and proliferation reduces the space of accessible configurations of generic polymer chains, in turn potentially affecting their mechanical~\cite{Patil2020} or biological~\cite{Postow2001,Bates2005,Sogo1999} functions. 
Entanglements are so inevitable and detrimental in the genome of living organisms that a specific class of highly conserved proteins -- known as topoisomerases -- has evolved to resolve them~\cite{Wang1985}. 

Statistical mechanical treatments of entanglements is difficult because these enter the problem as global constraints in the phase space of possible states that can be visited by the system. While it is typical to treat abundant topological constraints at the mean field level -- for instance in the tube model of polymer melts~\cite{Doi1988} -- exact and scaling results can be obtained via theories that replace entanglements with slip-links (SLs) which enforce contacts between polymer segments while allowing them to slide past each other~\cite{Ball1981,EdwardsVilgis1986,Higgs1989a,Michieletto2016softmatter}. Because of this physically appealing analogy, systems of polymers with slip-links have been theoretically and numerically explored in the field of statistical and polymer physics, for instance to estimate the size of knots~\cite{Metzler2002,Orlandini2009a} and the effective tube size in polymer melts~\cite{Likhtman2014a}. 

\begin{figure}[t!]
	\includegraphics[width=0.5\textwidth]{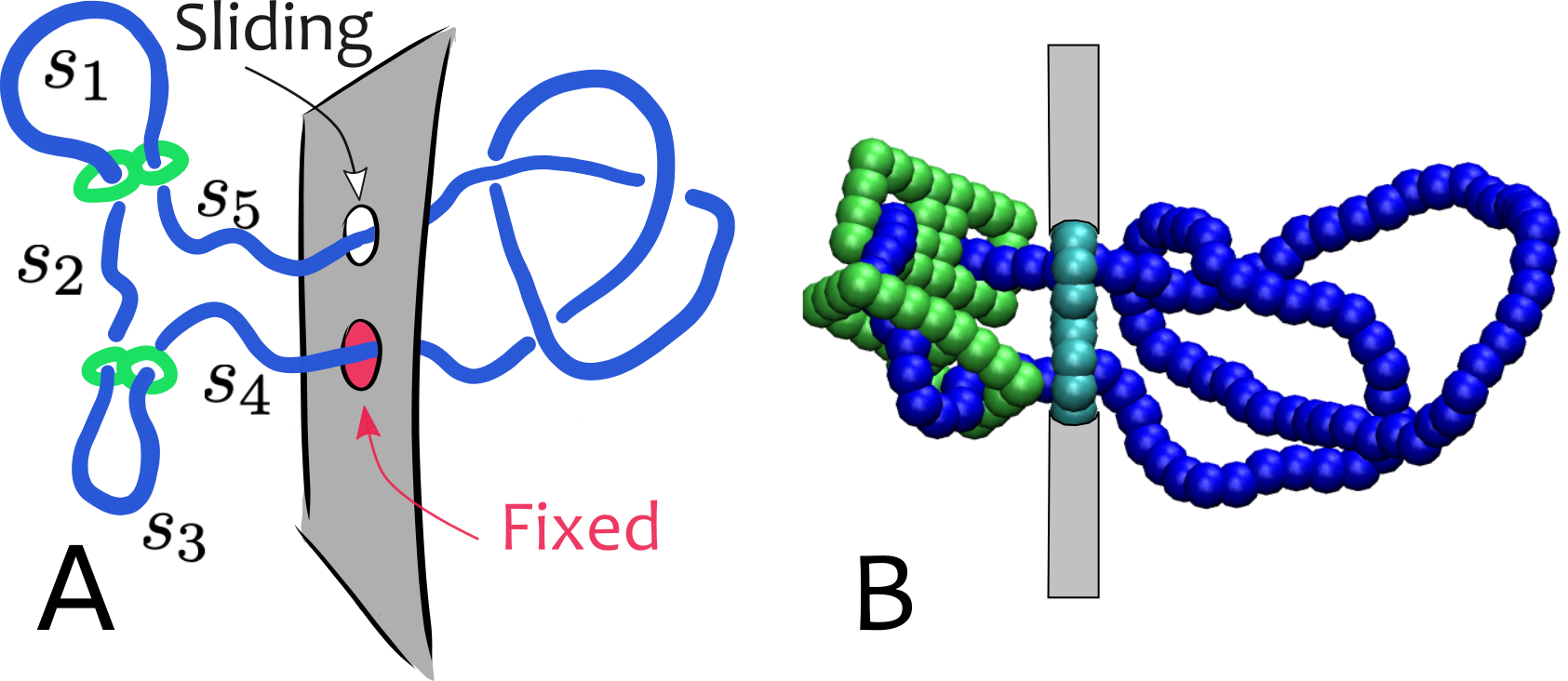}
	\vspace{-0.7 cm}
	\caption{\textbf{Entropic competition of SL and entanglements.} \textbf{A} Our set up is inspired to that of entropic competition~\cite{Metzler2002}: a ring polymer with total length $L$ is tied in a knot and contains diffusing SLs. The ring is forced to pass through two holes on a wall that separates the two sides. Only the contour passing through the top hole is allowed to slide, while the other is fixed. In the figure, the left-handed loop has $N_{SL}=2$ SLs and is partitioned into $2N_{SL}+1=5$ segments of which $s_1$ and $s_3$ are called peripheral, while $s_2$, $s_4$, and $s_5$ are called internal ($s_2+s_4+s_5$ makes up the inner loop). The right hand side contains a $3_1$ knot. \textbf{B} Snapshot from Molecular Dynamics simulations of the system in \textbf{A}.}
	\label{fig:sketch}
	\vspace{-0.5 cm}
\end{figure}

Beyond the theoretical appeal of these systems, actual examples in which SLs affect the conformation of polymers can be found in nature and industry. For instance, the family of Structural Maintenance of Chromosomes (SMC) proteins act as SLs on the genome and, depending on the condition and organism, they can either actively move~\cite{Fudenberg2016,Wang2017c,Ganji2018,Davidson2019a} or passively diffuse~\cite{Brackley2017prl,Davidson2016,Stigler2016,Ryu2020}, in turn extruding loops. These proteins are key for the organisation of the genome {\it in vivo}~\cite{Nasmyth2001,Hirano2002,Gibcus2018}, for the structure of sister chromatids~\cite{Goloborodko2016a,Gibcus2018} and to regulate chromatid decatenation and cell division~\cite{Piskadlo2017}. Additionally, it was recently suggested that they cooperate with topoisomerases to maintain the genome topologically simple and entanglement-free~\cite{Orlandini2019}.  While the structural role of SMC proteins is now widely explored, the mechanisms through which they regulate genome topology are far less understood or investigated. At the same time, SLs can be realised in synthetic and supra-molecular chemistry using cyclodextrins~\cite{Inoue2006} and are employed for instance to make molecular machines~\cite{Forgan2011} and slide-ring gels~\cite{Ito2007}. In these cases, the benefit of using SLs is that they effectively act as mobile cross-links, thus imposing strong, yet plastic, topological constraints on the polymeric constituents. Thanks to this peculiar feature, gels of polyrotaxanes display toughness and self-healing abilities far superior than those of traditionally cross-linked materials~\cite{BinImran2014}. As for biological systems, also the SL-mediated topological regulation of synthetic slide-ring gels is only starting to be investigated~\cite{Yasuda2019}.

In this work we study the interplay between SLs and topological entanglements formed by knots tied on a ring polymer (see Fig.~\ref{fig:sketch}). While our set up is inspired to the existing framework of entropic competition~\cite{Metzler2002,Zandi2003,Orlandini2009a}, here we consider SLs as real components of the system, modelling the presence of (diffusing) SMC proteins or cyclodextrins. 
In particular, we compute the loop size distribution and length of knotted segments for a variety of SL network topologies, and show that diffusing SLs are able to localise knotted segments purely by entropy, {\it without any external energy input}. We show that the efficiency of entanglement localisation depends on the particular topology of the SL network and that including the action of topoisomerases (modelled as transient strand-crossings) leads to extremely fast and efficient simplification of complex knots. Our results suggest an entropy-driven mechanism through which generic SL-like molecules can regulate the topology of DNA or synthetic polymers.

\paragraph*{Model -- }
We prepare the initial configuration of our system by joining two bead-spring polymer segments on either sides of a wall; typically, we consider configurations in which the two polymer segments are either knotted or contain SLs. Since in our model the polymer cannot cross the wall, itself or the SLs, the topology on each side is preserved throughout the simulations (see SM for further details of the Molecular Dynamics simulations and potentials used). Slip-links are modelled as physical square hand-cuffs and are allowed to slide diffusively along the polymer. The setup is such that the polymer passes through two holes on the wall, the size of which is small enough to let just one monomer through each of them at any time (see Fig.~\ref{fig:sketch}). Additionally, in line with previous works on entropic competition~\cite{Zandi2003}, we fix the position of one of the two beads closest to the wall so that the segments can exchange monomers only through the other hole (see Fig.~\ref{fig:sketch}). 

\begin{figure}[t!]
\includegraphics[width=0.45\textwidth]{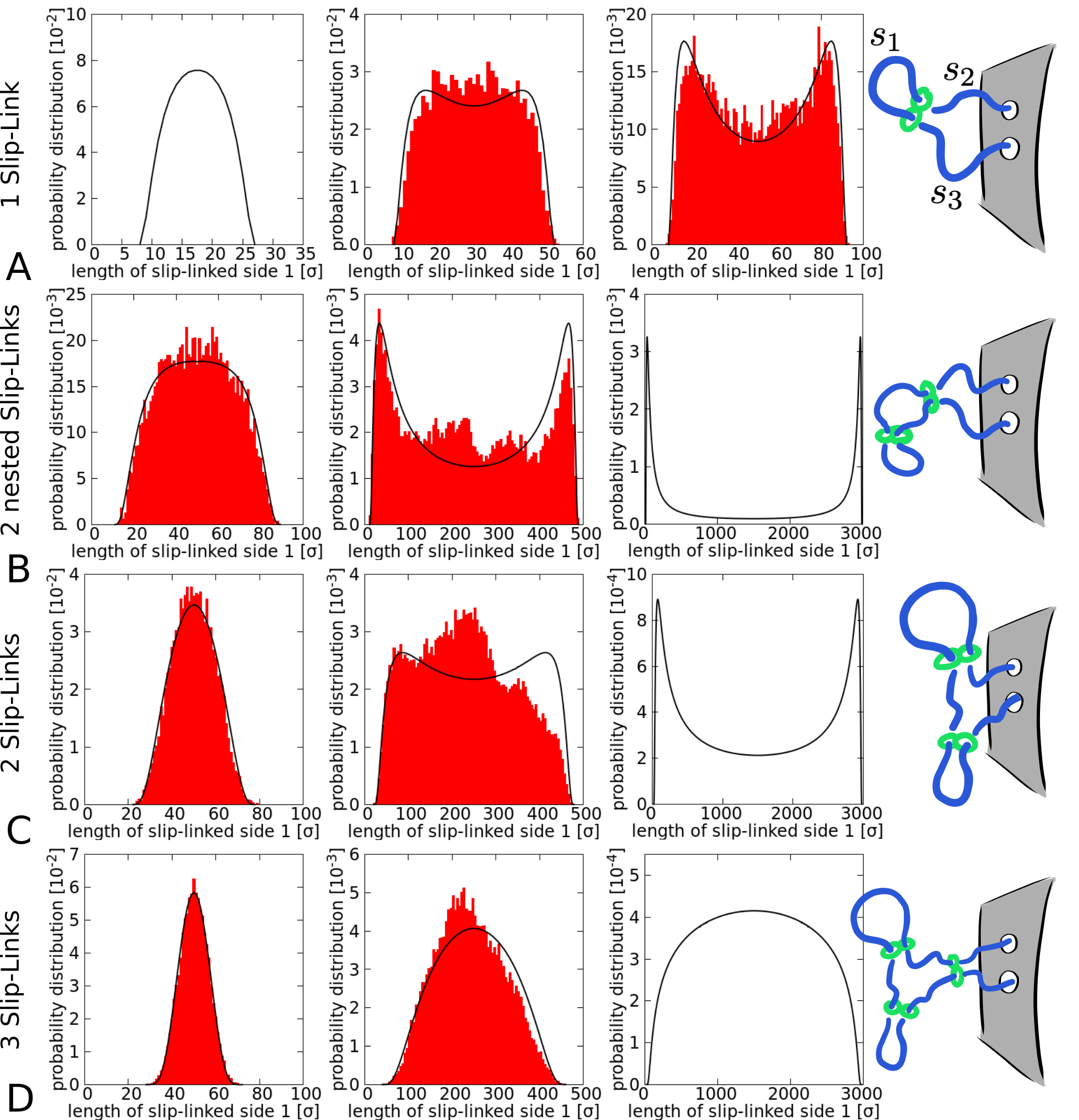}
\vspace{-0.2 cm}
\caption{\textbf{Entropic competition between symmetric SL networks}. From top to bottom: \textbf{A} a single SL; \textbf{B} $2$ SLs in a necklace configuration; \textbf{C} $2$ SLs in a round-table configuration; \textbf{D} $3$ SLs in a round-table configuration. Each histogram collects information from $10^4$ configurations sampled every $10^6$ LAMMPS steps from $100$ independent replicas. The black curves are obtained solving Eq.~\eqref{eq:scalingRT} for the relevant SL network topology. From left to right: increasing total length of the ring polymer (35, 60, 100 beads in \textbf{A}) and (100, 500, 3000 beads) in \textbf{B-D}).}
\vspace{-0.4 cm}
\label{fig:SL_vs_SL}
\end{figure}

\paragraph*{Competition of SLs -- }
We first study the competition between symmetric SLs networks: 
we load up to $3$ SLs on each side of the polymer in either a ``round table'' (RT) configuration, realised by loading the slip-links in series~\cite{Metzler2002} (Fig.~\ref{fig:SL_vs_SL}A,C,D), or a ``necklace'' (NC) configuration, characterised by nested loops~\cite{Metzler2002} (Fig.~\ref{fig:SL_vs_SL}B). The number of configurations of a ring of length $L$ containing $N_{SL}$ slip-links in the RT configuration and grafted to an impenetrable surface as in Fig.~\ref{fig:SL_vs_SL} scales as (see SM)
\begin{align}
&\mathcal{Z}_{RT,N_{SL}}(L) \sim \int{ds_1}\int{ds_2}\dots\int{ds_{2N_{SL}+1}} \times \notag \\
&\times \left[\mu^L\left(L-\sum^{N_{SL}}_{i=1}{s_i} \right)^{-3\nu-1}\prod^{N_{SL}}_{i=1}{s_i}^{\alpha}\right]_{\sum_{i=1}^{2N_{SL}+1}{s_i}=L},
\label{eq:scalingRT}
\end{align}
where $s_i$, $i=1,\dots,N_{SL}$ are the lengths of the $N_{SL}$ peripheral loops, $s_i$, $i=N_{SL}+1,\dots,2N_{SL}+1$ are the lengths of the other $N_{SL}+1$ segments (see Fig.~\ref{fig:sketch}, note that the wall is not considered as a SL), $\nu$ is the universal metric exponent relating the size of a polymer to its length and $\mu$ is the non-universal connectivity constant. By using known results for the statistical physics of polymer networks~\cite{Duplantier1989,Metzler2002}, we can convert a polymer containing $N_{SL}$ SLs into a network of $N_{SL}+1$ loops with $2N_{SL}+1$ edges and $N_{SL}$ $4$-legged nodes (see Fig.~\ref{fig:sketch}A). Within this framework, $\alpha$ can be computed as
\begin{equation}
\label{eq:alpha}
\alpha=-3\nu+\sigma_{4},
\end{equation}
with $\sigma_{4}=-0.46$ the scaling exponent associated with $4$-legged nodes~\cite{Duplantier1989}. 
The integration extends over all lengths compatible with the given topology, and accounts for the sliding entropy of the SLs. We note that Eq.~\eqref{eq:scalingRT} assumes that the peripheral loops are tight with respect to the inner one (this assumption needs to be verified self-consistently a posteriori). Since the two competing sides of the polymer are not interacting, the total number of configurations can be factorised as $\mathcal{Z}_{RT,N_1}(L_1)\times\mathcal{Z}_{RT,N_2}(L_2)$. 

\begin{figure*}[t!]
	\includegraphics[width=1\textwidth]{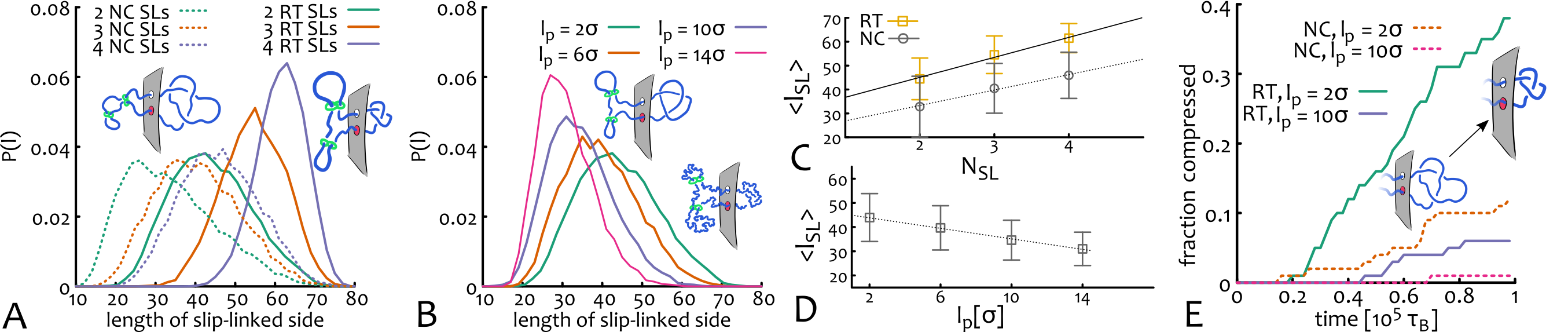}
	\vspace{-0.5 cm}
	\caption{\textbf{Entropic Competition between SLs and a Trefoil Knot.} \textbf{A} Plot of the distribution of lengths of the side with the SLs, for different SL number and topology. The plot shows that the RT configurations (solid lines) favour longer SL sides and hence tighter knots than NC configurations (dashed lines). \textbf{B} Distribution of lengths of the side with the SLs, for different substrate persistence lengths: the curves refer to the case of a RT configuration with $2$ SLs. The plot shows that stiffer substrates favour longer knotted sides. \textbf{C} Mean length ($\langle l_{SL} \rangle$) and standard error of the SL side length as a function of SL number and topology. The solid and dashed lines are best linear fits with slopes $8.4(8)$ for RT and $6.5(6)$ for NC. \textbf{D} Mean length ($\langle l_{SL} \rangle$) and standard error of the SL side length as a function of substrate persistence length $l_p$. The solid line is the best linear fit with slope $- 1.09(5) l_p$. \textbf{E}  Fraction of knots that have been compressed -- i.e., confined within $20\%$ of the overall ring contour length -- during the course of a simulation starting from a situation in which $80\%$ of the contour length is on the knotted side. Higher flexibility and RT configurations both favour knot localisation. NC=necklace; RT=round-table; SLs=slip-links.}
	\label{fig:SL_vs_knot}
	\vspace{-0.5 cm}
\end{figure*}

We employ Eq.~\eqref{eq:scalingRT} to derive semi-analytically the distribution of loop lengths for systems with up to $3$ SLs on each side. We then verify these predictions with Molecular Dynamics (MD) simulations and report the results in Fig.~\ref{fig:SL_vs_SL}. 
Figure~\ref{fig:SL_vs_SL}A shows that the distribution undergoes a unimodal-bimodal transition for increasing ring sizes (from left to right we consider 35, 60 and 100 beads-long polymers). For short chains the segments on either side are comparable, whereas for longer ones there is symmetry breaking with one of the two sides taking up most of the contour length. The transition is associated with a pitchfork bifurcation and occurs at the critical length $L_c \simeq 50$ for this set up with 1 SL (see SM).

We now turn to the case of $2$ and $3$ SLs on each side (Fig.~\ref{fig:SL_vs_SL}B-D). Since $2$ SLs in a NC configuration can also be viewed as a RT, the distribution of contour lengths can be computed using Eq.~\eqref{eq:scalingRT} for both these topologies (note that the identification of the peripheral loops varies between NC and RT arrangements with $2$ SLs, see SM). We find that while the NC topology yields symmetry breaking for long enough chains, the RT one displays a stable symmetric state for all polymer lengths when $N_{SL}=3$ SL are loaded on each side (Fig.~\ref{fig:SL_vs_SL}D).
In light of these results we argue that the network topology arising from the loading of multiple SLs -- either in series (RT) or in parallel (NC) -- profoundly affects the spatial organisation of the underlying DNA or synthetic polymer. Further, our semi-analytical results are in line with previous works which showed that nested SLs (NC topology) promote entropic ratcheting of the outer loop so that longer DNA loops are extruded on average~\cite{Brackley2017prl}. This corresponds to the case in which there is symmetry breaking and one of the loops grows extensively with the total contour length (Fig.~\ref{fig:SL_vs_SL}B,C).

\paragraph{SLs versus knots -- } 
The results of the previous section suggest that entropic competition of SLs can regulate the distribution of loop lengths of an underlying polymer. 
We now ask whether this entropic competition may also provide a mechanism for the topological simplification of entanglements such as knots that may occur on biological or synthetic polymers. To address this question, we now consider the case where the competing topologies are a SL network and a trefoil ($3_1$) knot (as in Fig.~\ref{fig:sketch}), and aim to identify the optimal conditions to simplify the knotted loop. 
It should be noted that our calculations are performed in thermal equilibrium, so we seek entropic and topological mechanisms that bias the free energy towards a lower knotting probability. This is different from previous works which studied the effects of energy-consuming mechanisms driving unknotting~\cite{Orlandini2019}.

Since the semi-analytical approach we used in the previous sections cannot be easily extended to this setup~\cite{noteknot}, we directly perform MD simulations of a $100$ beads-long ring and address how different factors such as SL number, topology and polymer persistence length affect the entropic competition and, in turn, regulate the probability and efficiency of knot localisation. 

First, we find a large difference between RT versus NC configurations (Fig.~\ref{fig:SL_vs_knot}A): the larger sliding entropy and low looping cost of the RT set-up provides an entropic pressure which outcompetes the ratcheting effect of the NC topology. This result is unexpected, since it is known that NC configurations promote asymmetric polymer conformations and the formation of large loops with respect to RT networks (see Fig.~\ref{fig:SL_vs_SL}B,C and \cite{Brackley2017prl}). Nevertheless, the RT arrangement appears to be more suited at overcoming the knot entropy. Quantitatively, the entropic pressure of the RT configurations is notable: already with $3$ SLs the knotted side is localised to a shorter contour length than the SL side (Fig.~\ref{fig:SL_vs_knot}A).  

Additionally, our simulations reveal that the bending rigidity of the polymer substrate also plays an important role (Fig.~\ref{fig:SL_vs_knot}B). Indeed, there are two potential and contrasting enthalpic effects that the bending rigidity may have on the entropic competition: on the one hand, the larger the stiffness, the longer the contour length required to form a knot without tight bends (this effect is particularly relevant for short loops); on the other hand, bending rigidity enhances the formation of large loops via diffusive loop extrusion~\cite{Bonato2020}. Our simulations show that -- for $2$ SLs in a RT arrangement and a $100$-bead polymer ring -- the former effect dominates: the larger the persistence length, the larger the knotted side, rendering the localisation of entanglements less efficient (Fig.~\ref{fig:SL_vs_knot}B). This result shows that the entropic pressure provided by the SLs can only provide localisation if the polymer substrate is sufficiently flexible. 

More quantitatively, in Fig.~\ref{fig:SL_vs_knot}C we show that the mean length of the SL-side, $\langle l_{SL} \rangle$, grows linearly as a function of the number of SLs (in the range of $N_{SL}$ explored here) and that the growth rate is faster for RT configurations. At the same time, $\langle l_{SL} \rangle$ appears to also decrease linearly with polymer stiffness as $\sim - l_p$ (Fig.~\ref{fig:SL_vs_knot}D). In a practical genomic or synthetic context, the results reported in Fig.~\ref{fig:SL_vs_knot}A-D overall suggest that local SL density and substrate flexibility may provide handles to tune the typical size of knots. More specifically, in vivo both these parameters depend on local DNA sequence and transcriptional activity, and we thus speculate that knot and entanglement localisation may be achieved to a different extent in different genomic regions; in particular, transcriptionally active genes (which are also thought to be more flexible) may harbour smaller knots and localised entanglements.  At the same time, different synthetic polymers 
have different stiffness and different arrangements of cyclodextrins may be designed, so that the resulting cross-linked gels have tunable properties~\cite{BinImran2014}.
 
 
\paragraph{Kinetics of Topological Simplification -- } 
Up to now we have used the framework of entropic competition to study knot localisation in an equilibrium framework. Another important aspect of the problem is {\it how fast} knots can be localised, and eventually removed. 

To measure the kinetics of knot localisation, we perform MD simulations in which the system is initialised far from equilibrium with a large knot -- occupying $80\%$ of the total contour length -- and measure how long it takes for the system to revert the situation and compress the knotted size to $20\%$ of the total contour length. Figure~\ref{fig:SL_vs_knot}E shows that RT configurations and flexible substrates are faster at localising the knot, on top of being more efficient in steady state, as discussed before.

\begin{figure}[t!]
	\includegraphics[width=0.5\textwidth]{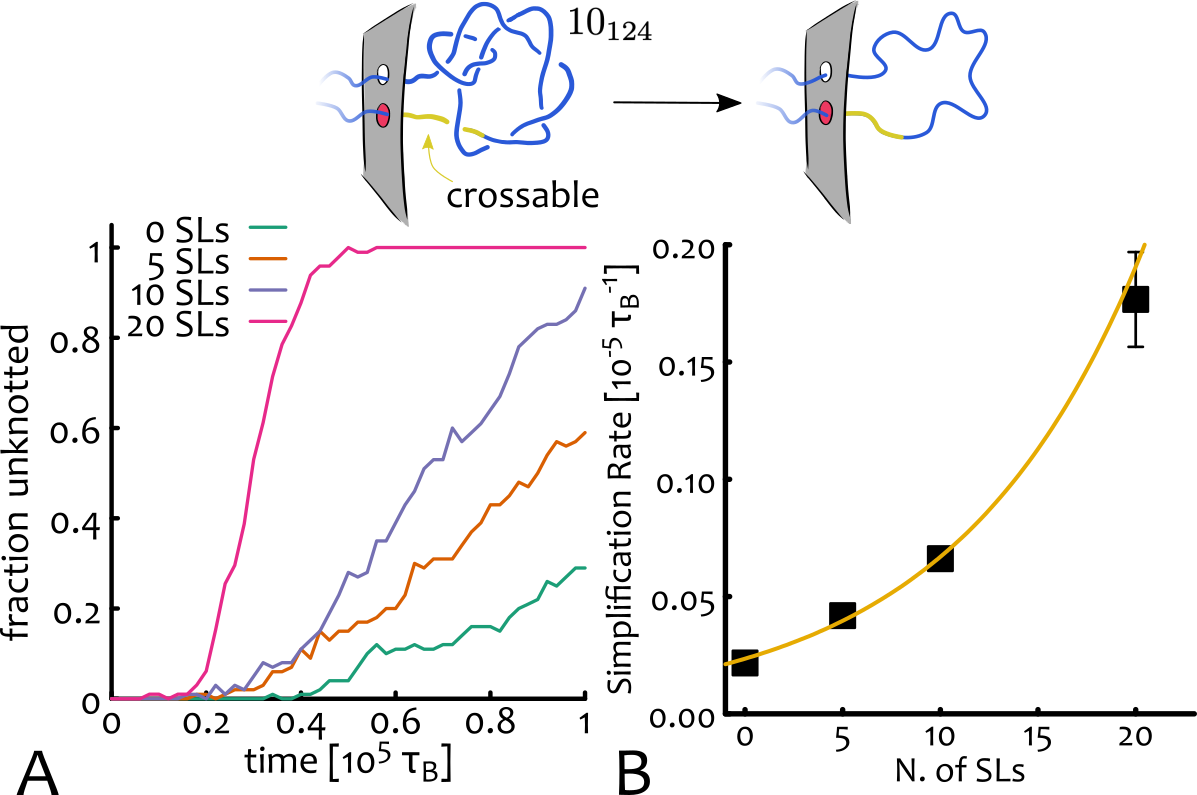}
	\vspace{-0.3 cm}
	\caption{\textbf{Kinetics of Topological Simplification.}  \textbf{A} Fraction of $10_{124}$ knots that have simplified to unknots during the course of simulations of a polymer with $N=400$ beads. starting from a situation in which $80\%$ of the contour length is on the knotted side. In this simulation, soft interactions between the $10$ closest beads to the bead fixed at the wall on the knotted side are used to allow occasional strand-crossings to model topoisomerase action close to a SL. Loading more SLs increases the pressure on the knot and hence the rate of unknotting. \textbf{B} The simplification rate (measured as the slope of the curves in \textbf{A}) shows an exponential increase as a function of number of SLs. SL=slip-links.}
	\label{fig:SL_vs_knot_kinetics}
	\vspace{-0.5 cm}
\end{figure}

Finally, to address the question of whether diffusing SLs can drive the topological simplification of knots and entanglements we now include topoisomerase-mediated strand-crossing reactions in our model as follows. 
We study a system in which a short ($10$ beads) segment of the knotted side is allowed to undergo strand-crossings (see SM). We then initialise the system with a $10$-crossings torus knot ($10_{124}$) taking up $75\%$ of the total contour length on the right-hand side and load $N=\{0,5,10,20\}$ SLs in RT arrangements on the other side. We also choose to set only the $10$ closest beads to the wall as crossable, modelling the presence of a topoisomerase closely upstream of a SL-like SMC protein, as suggested by recent experimental evidence~\cite{Vian2018a,Uuskula-Reimand2016} (note that in this context we view the wall in Fig.~\ref{fig:sketch} as a SL itself).
	
With this set-up, we observe that the more SLs are loaded onto the substrate, the faster the rate of the topological simplification (see Fig.~\ref{fig:SL_vs_knot_kinetics}A). The quantitative speed-up is striking, as the rate grows exponentially with $N_{SL}$. An explanation of this dependence is that the mean length of the SL side scales linearly with the number of SLs (Fig.~\ref{fig:SL_vs_knot}C), so that the mean knotted length is $\langle l_{\mathcal{K}} \rangle = L - \langle l_{SL} \rangle \simeq L - k N_{SL}$. Since the probability of observing a trivial knot on a polymer $x$ segments long scales as $P_{0}(L) \sim e^{-x/x_0}$~\cite{Pippenger1989,Micheletti2006c}, with $x_0$ a model-specific parameter, we expect $P_{0}(\langle l_{\mathcal{K}} \rangle) \sim e^{-\langle l_{\mathcal{K}} \rangle/L_0} \sim e^{N_{SL}/L_0}$. In other words, as the knot size becomes smaller, its entropic cost increases sharply and it becomes exponentially harder to prevent its simplification to the unknot. [Note that this argument implicitly assumes that simplification is slow with respect to localisation.]

\paragraph{Conclusions -- }
In summary, in the context of a search for the possible mechanisms for topological regulation and simplification of entanglements in DNA and other polymeric systems, here we have investigated the interplay between the entropy of SL networks and of knotted topologies.

In the case where no knots are tied on the ring, (SL-only case, Fig.~\ref{fig:SL_vs_SL}), we provide semi-analytical predictions on the distribution of polymer lengths which compare well to direct MD simulations. We then consider a set-up in which an SL network competes entropically with a trefoil knot and we find that the network can localise the knot efficiently, by just relying on its higher entropic pressure and in the absence of any motor activity associated with the SLs (such an activity is known to be absent in yeast cohesin~\cite{Ryu2020} and cyclodextrins~\cite{BinImran2014}). We dissect the effects of the number of SLs, the topology of the SL network, and the polymer flexibility on the knot localisation discovering that round-table (or ``in series'') SL arrangements are best suited at confining a knot. Additionally, when the substrate is stiff, the enthalpic contribution from the bending energy dominates and the knot swells, hampering localisation. Finally, we have also included a topoisomerase activity in the model, which is relevant to study topological simplification in the genomes of living cells~\cite{Orlandini2019} or self-healing gels of polyrotaxanes with reversible bonds~\cite{Nakahata2016}. Strikingly, we find that entropic pressure alone is sufficient to simplify very complicated knots reliably and extremely fast, with the simplification rate increasing exponentially with number of SLs. 

Our findings inform the design of strategies for regulating, simplifying or preserving desired topologies in synthetic polymers, for instance by targeted design of SL arrangement and density or substrate stiffness. Additionally, they also help to explain the mechanisms of topological simplification by SL-like proteins such as SMC. 

\paragraph{Acknowledgements--}
DMi is supported by the Leverhulme Trust through an Early Career Fellowship (ECF-2019-088). 
 
\bibliographystyle{apsrev4-1}
\bibliography{library}

\begin{thebibliography}{45}%
\makeatletter
\providecommand \@ifxundefined [1]{%
 \@ifx{#1\undefined}
}%
\providecommand \@ifnum [1]{%
 \ifnum #1\expandafter \@firstoftwo
 \else \expandafter \@secondoftwo
 \fi
}%
\providecommand \@ifx [1]{%
 \ifx #1\expandafter \@firstoftwo
 \else \expandafter \@secondoftwo
 \fi
}%
\providecommand \natexlab [1]{#1}%
\providecommand \enquote  [1]{``#1''}%
\providecommand \bibnamefont  [1]{#1}%
\providecommand \bibfnamefont [1]{#1}%
\providecommand \citenamefont [1]{#1}%
\providecommand \href@noop [0]{\@secondoftwo}%
\providecommand \href [0]{\begingroup \@sanitize@url \@href}%
\providecommand \@href[1]{\@@startlink{#1}\@@href}%
\providecommand \@@href[1]{\endgroup#1\@@endlink}%
\providecommand \@sanitize@url [0]{\catcode `\\12\catcode `\$12\catcode
  `\&12\catcode `\#12\catcode `\^12\catcode `\_12\catcode `\%12\relax}%
\providecommand \@@startlink[1]{}%
\providecommand \@@endlink[0]{}%
\providecommand \url  [0]{\begingroup\@sanitize@url \@url }%
\providecommand \@url [1]{\endgroup\@href {#1}{\urlprefix }}%
\providecommand \urlprefix  [0]{URL }%
\providecommand \Eprint [0]{\href }%
\providecommand \doibase [0]{http://dx.doi.org/}%
\providecommand \selectlanguage [0]{\@gobble}%
\providecommand \bibinfo  [0]{\@secondoftwo}%
\providecommand \bibfield  [0]{\@secondoftwo}%
\providecommand \translation [1]{[#1]}%
\providecommand \BibitemOpen [0]{}%
\providecommand \bibitemStop [0]{}%
\providecommand \bibitemNoStop [0]{.\EOS\space}%
\providecommand \EOS [0]{\spacefactor3000\relax}%
\providecommand \BibitemShut  [1]{\csname bibitem#1\endcsname}%
\let\auto@bib@innerbib\@empty
\bibitem [{\citenamefont {C}\ \emph {et~al.}(1996)\citenamefont {C},
  \citenamefont {De~Griend~Pieter},\ and\ \citenamefont
  {Charles}}]{history_science_knots}%
  \BibitemOpen
  \bibfield  {author} {\bibinfo {author} {\bibfnamefont {T.}~\bibnamefont {C}},
  \bibinfo {author} {\bibfnamefont {V.}~\bibnamefont {De~Griend~Pieter}}, \
  and\ \bibinfo {author} {\bibfnamefont {W.}~\bibnamefont {Charles}},\ }\href
  {https://books.google.co.uk/books?id=ovDsCgAAQBAJ} {\emph {\bibinfo {title}
  {History And Science Of Knots}}},\ Series On Knots And Everything\ (\bibinfo
  {publisher} {World Scientific Publishing Company},\ \bibinfo {year}
  {1996})\BibitemShut {NoStop}%
\bibitem [{\citenamefont {van Rijssel}\ \emph {et~al.}(1990)\citenamefont {van
  Rijssel}, \citenamefont {Trimbos},\ and\ \citenamefont
  {Booster}}]{VanRijssel1990}%
  \BibitemOpen
  \bibfield  {author} {\bibinfo {author} {\bibfnamefont {E.~J.}\ \bibnamefont
  {van Rijssel}}, \bibinfo {author} {\bibfnamefont {J.~B.}\ \bibnamefont
  {Trimbos}}, \ and\ \bibinfo {author} {\bibfnamefont {M.~H.}\ \bibnamefont
  {Booster}},\ }\href {\doibase https://doi.org/10.1016/0002-9378(90)90828-U}
  {\bibfield  {journal} {\bibinfo  {journal} {American Journal of Obstetrics
  and Gynecology}\ }\textbf {\bibinfo {volume} {162}},\ \bibinfo {pages} {93 }
  (\bibinfo {year} {1990})}\BibitemShut {NoStop}%
\bibitem [{\citenamefont {Chen}\ \emph {et~al.}(1995)\citenamefont {Chen},
  \citenamefont {Rauch}, \citenamefont {White}, \citenamefont {Englund},\ and\
  \citenamefont {Cozzarelli}}]{Chen1995}%
  \BibitemOpen
  \bibfield  {author} {\bibinfo {author} {\bibfnamefont {J.}~\bibnamefont
  {Chen}}, \bibinfo {author} {\bibfnamefont {C.~A.}\ \bibnamefont {Rauch}},
  \bibinfo {author} {\bibfnamefont {J.~H.}\ \bibnamefont {White}}, \bibinfo
  {author} {\bibfnamefont {P.~T.}\ \bibnamefont {Englund}}, \ and\ \bibinfo
  {author} {\bibfnamefont {N.}~\bibnamefont {Cozzarelli}},\ }\href
  {http://www.ncbi.nlm.nih.gov/pubmed/7813018} {\bibfield  {journal} {\bibinfo
  {journal} {Cell}\ }\textbf {\bibinfo {volume} {80}},\ \bibinfo {pages} {61}
  (\bibinfo {year} {1995})}\BibitemShut {NoStop}%
\bibitem [{\citenamefont {Wasserman}\ \emph {et~al.}(1985)\citenamefont
  {Wasserman}, \citenamefont {Dungan},\ and\ \citenamefont
  {Cozzarelli}}]{Wasserman1985a}%
  \BibitemOpen
  \bibfield  {author} {\bibinfo {author} {\bibfnamefont {S.}~\bibnamefont
  {Wasserman}}, \bibinfo {author} {\bibfnamefont {J.}~\bibnamefont {Dungan}}, \
  and\ \bibinfo {author} {\bibfnamefont {N.}~\bibnamefont {Cozzarelli}},\
  }\href {\doibase 10.1126/science.2990045} {\bibfield  {journal} {\bibinfo
  {journal} {Science (80-. ).}\ }\textbf {\bibinfo {volume} {229}},\ \bibinfo
  {pages} {171} (\bibinfo {year} {1985})}\BibitemShut {NoStop}%
\bibitem [{\citenamefont {Patil}\ \emph {et~al.}(2020)\citenamefont {Patil},
  \citenamefont {Sandt}, \citenamefont {Kolle},\ and\ \citenamefont
  {Dunkel}}]{Patil2020}%
  \BibitemOpen
  \bibfield  {author} {\bibinfo {author} {\bibfnamefont {V.~P.}\ \bibnamefont
  {Patil}}, \bibinfo {author} {\bibfnamefont {J.~D.}\ \bibnamefont {Sandt}},
  \bibinfo {author} {\bibfnamefont {M.}~\bibnamefont {Kolle}}, \ and\ \bibinfo
  {author} {\bibfnamefont {J.}~\bibnamefont {Dunkel}},\ }\href {\doibase
  10.1126/science.aaz0135} {\bibfield  {journal} {\bibinfo  {journal}
  {Science}\ }\textbf {\bibinfo {volume} {367}},\ \bibinfo {pages} {71}
  (\bibinfo {year} {2020})}\BibitemShut {NoStop}%
\bibitem [{\citenamefont {Postow}\ \emph {et~al.}(2001)\citenamefont {Postow},
  \citenamefont {Crisona}, \citenamefont {Peter}, \citenamefont {Hardy},\ and\
  \citenamefont {Cozzarelli}}]{Postow2001}%
  \BibitemOpen
  \bibfield  {author} {\bibinfo {author} {\bibfnamefont {L.}~\bibnamefont
  {Postow}}, \bibinfo {author} {\bibfnamefont {N.~J.}\ \bibnamefont {Crisona}},
  \bibinfo {author} {\bibfnamefont {B.~J.}\ \bibnamefont {Peter}}, \bibinfo
  {author} {\bibfnamefont {C.~D.}\ \bibnamefont {Hardy}}, \ and\ \bibinfo
  {author} {\bibfnamefont {N.~R.}\ \bibnamefont {Cozzarelli}},\ }\href
  {\doibase 10.1073/pnas.111006998} {\bibfield  {journal} {\bibinfo  {journal}
  {Proc. Nat. Acad. Sci. USA}\ }\textbf {\bibinfo {volume} {98}},\ \bibinfo
  {pages} {8219} (\bibinfo {year} {2001})}\BibitemShut {NoStop}%
\bibitem [{\citenamefont {Bates}\ and\ \citenamefont
  {Maxwell}(2005)}]{Bates2005}%
  \BibitemOpen
  \bibfield  {author} {\bibinfo {author} {\bibfnamefont {A.}~\bibnamefont
  {Bates}}\ and\ \bibinfo {author} {\bibfnamefont {A.}~\bibnamefont
  {Maxwell}},\ }\href
  {http://books.google.com/books?hl=en{\&}lr={\&}id=WGBAGyzvQOUC{\&}oi=fnd{\&}pg=PR17{\&}dq=DNA+topology{\&}ots=TUaM8kASav{\&}sig=56tWxeOcV-zwI3l9c030FtWj1Y0}
  {\emph {\bibinfo {title} {{DNA topology}}}}\ (\bibinfo  {publisher} {Oxford
  University Press},\ \bibinfo {year} {2005})\BibitemShut {NoStop}%
\bibitem [{\citenamefont {Sogo}\ \emph {et~al.}(1999)\citenamefont {Sogo},
  \citenamefont {Stasiak}, \citenamefont {Mart{\'{i}}nez-Robles}, \citenamefont
  {Krimer}, \citenamefont {Hern{\'{a}}ndez},\ and\ \citenamefont
  {Schvartzman}}]{Sogo1999}%
  \BibitemOpen
  \bibfield  {author} {\bibinfo {author} {\bibfnamefont {J.~M.}\ \bibnamefont
  {Sogo}}, \bibinfo {author} {\bibfnamefont {A.}~\bibnamefont {Stasiak}},
  \bibinfo {author} {\bibfnamefont {M.~L.}\ \bibnamefont
  {Mart{\'{i}}nez-Robles}}, \bibinfo {author} {\bibfnamefont {D.~B.}\
  \bibnamefont {Krimer}}, \bibinfo {author} {\bibfnamefont {P.}~\bibnamefont
  {Hern{\'{a}}ndez}}, \ and\ \bibinfo {author} {\bibfnamefont {J.~B.}\
  \bibnamefont {Schvartzman}},\ }\href
  {http://www.ncbi.nlm.nih.gov/pubmed/10024438} {\bibfield  {journal} {\bibinfo
   {journal} {J. Mol. Biol.}\ }\textbf {\bibinfo {volume} {286}},\ \bibinfo
  {pages} {637} (\bibinfo {year} {1999})}\BibitemShut {NoStop}%
\bibitem [{\citenamefont {Wang}(1985)}]{Wang1985}%
  \BibitemOpen
  \bibfield  {author} {\bibinfo {author} {\bibfnamefont {J.~C.}\ \bibnamefont
  {Wang}},\ }\href {\doibase 10.1146/annurev.bi.54.070185.003313} {\bibfield
  {journal} {\bibinfo  {journal} {Annu. Rev. Biochem.}\ }\textbf {\bibinfo
  {volume} {54}},\ \bibinfo {pages} {665} (\bibinfo {year} {1985})}\BibitemShut
  {NoStop}%
\bibitem [{\citenamefont {Doi}\ and\ \citenamefont {Edwards}(1988)}]{Doi1988}%
  \BibitemOpen
  \bibfield  {author} {\bibinfo {author} {\bibfnamefont {M.}~\bibnamefont
  {Doi}}\ and\ \bibinfo {author} {\bibfnamefont {S.}~\bibnamefont {Edwards}},\
  }\href@noop {} {\emph {\bibinfo {title} {{The theory of polymer dynamics}}}}\
  (\bibinfo  {publisher} {Oxford University Press},\ \bibinfo {year}
  {1988})\BibitemShut {NoStop}%
\bibitem [{\citenamefont {Ball}\ \emph {et~al.}(1981)\citenamefont {Ball},
  \citenamefont {Doi}, \citenamefont {Edwards},\ and\ \citenamefont
  {Warner}}]{Ball1981}%
  \BibitemOpen
  \bibfield  {author} {\bibinfo {author} {\bibfnamefont {R.~C.}\ \bibnamefont
  {Ball}}, \bibinfo {author} {\bibfnamefont {M.}~\bibnamefont {Doi}}, \bibinfo
  {author} {\bibfnamefont {S.~F.}\ \bibnamefont {Edwards}}, \ and\ \bibinfo
  {author} {\bibfnamefont {M.}~\bibnamefont {Warner}},\ }\href@noop {}
  {\bibfield  {journal} {\bibinfo  {journal} {Polymer (Guildf).}\ }\textbf
  {\bibinfo {volume} {22}},\ \bibinfo {pages} {1010} (\bibinfo {year}
  {1981})}\BibitemShut {NoStop}%
\bibitem [{\citenamefont {Edwards}\ and\ \citenamefont
  {Vilgis}(1986)}]{EdwardsVilgis1986}%
  \BibitemOpen
  \bibfield  {author} {\bibinfo {author} {\bibfnamefont {S.}~\bibnamefont
  {Edwards}}\ and\ \bibinfo {author} {\bibfnamefont {T.}~\bibnamefont
  {Vilgis}},\ }\href@noop {} {\bibfield  {journal} {\bibinfo  {journal}
  {Polymer}\ }\textbf {\bibinfo {volume} {27}},\ \bibinfo {pages} {483}
  (\bibinfo {year} {1986})}\BibitemShut {NoStop}%
\bibitem [{\citenamefont {Higgs}\ and\ \citenamefont
  {Ball}(1989)}]{Higgs1989a}%
  \BibitemOpen
  \bibfield  {author} {\bibinfo {author} {\bibfnamefont {P.~G.}\ \bibnamefont
  {Higgs}}\ and\ \bibinfo {author} {\bibfnamefont {R.~C.}\ \bibnamefont
  {Ball}},\ }\href {\doibase 10.1209/0295-5075/8/4/010} {\bibfield  {journal}
  {\bibinfo  {journal} {Epl}\ }\textbf {\bibinfo {volume} {8}},\ \bibinfo
  {pages} {357} (\bibinfo {year} {1989})}\BibitemShut {NoStop}%
\bibitem [{\citenamefont {Michieletto}(2016)}]{Michieletto2016softmatter}%
  \BibitemOpen
  \bibfield  {author} {\bibinfo {author} {\bibfnamefont {D.}~\bibnamefont
  {Michieletto}},\ }\href {\doibase 10.1039/C6SM02168A} {\bibfield  {journal}
  {\bibinfo  {journal} {Soft Matter}\ }\textbf {\bibinfo {volume} {12}},\
  \bibinfo {pages} {9485} (\bibinfo {year} {2016})}\BibitemShut {NoStop}%
\bibitem [{\citenamefont {Metzler}\ \emph {et~al.}(2002)\citenamefont
  {Metzler}, \citenamefont {Hanke}, \citenamefont {Dommersnes}, \citenamefont
  {Kantor},\ and\ \citenamefont {Kardar}}]{Metzler2002}%
  \BibitemOpen
  \bibfield  {author} {\bibinfo {author} {\bibfnamefont {R.}~\bibnamefont
  {Metzler}}, \bibinfo {author} {\bibfnamefont {A.}~\bibnamefont {Hanke}},
  \bibinfo {author} {\bibfnamefont {P.~G.}\ \bibnamefont {Dommersnes}},
  \bibinfo {author} {\bibfnamefont {Y.}~\bibnamefont {Kantor}}, \ and\ \bibinfo
  {author} {\bibfnamefont {M.}~\bibnamefont {Kardar}},\ }\href {\doibase
  10.1103/PhysRevE.65.061103} {\bibfield  {journal} {\bibinfo  {journal} {Phys.
  Rev. E}\ }\textbf {\bibinfo {volume} {65}},\ \bibinfo {pages} {1} (\bibinfo
  {year} {2002})}\BibitemShut {NoStop}%
\bibitem [{\citenamefont {Orlandini}\ \emph {et~al.}(2009)\citenamefont
  {Orlandini}, \citenamefont {Stella},\ and\ \citenamefont
  {Vanderzande}}]{Orlandini2009a}%
  \BibitemOpen
  \bibfield  {author} {\bibinfo {author} {\bibfnamefont {E.}~\bibnamefont
  {Orlandini}}, \bibinfo {author} {\bibfnamefont {A.~L.}\ \bibnamefont
  {Stella}}, \ and\ \bibinfo {author} {\bibfnamefont {C.}~\bibnamefont
  {Vanderzande}},\ }\href {\doibase 10.1088/1478-3975/6/2/025012} {\bibfield
  {journal} {\bibinfo  {journal} {Phys. Biol.}\ }\textbf {\bibinfo {volume}
  {6}},\ \bibinfo {pages} {025012} (\bibinfo {year} {2009})}\BibitemShut
  {NoStop}%
\bibitem [{\citenamefont {Likhtman}(2014)}]{Likhtman2014a}%
  \BibitemOpen
  \bibfield  {author} {\bibinfo {author} {\bibfnamefont {A.~E.}\ \bibnamefont
  {Likhtman}},\ }\href {\doibase 10.1039/c3sm52575a} {\bibfield  {journal}
  {\bibinfo  {journal} {Soft Matter}\ }\textbf {\bibinfo {volume} {10}},\
  \bibinfo {pages} {1895} (\bibinfo {year} {2014})}\BibitemShut {NoStop}%
\bibitem [{\citenamefont {Fudenberg}\ \emph {et~al.}(2016)\citenamefont
  {Fudenberg}, \citenamefont {Imakaev}, \citenamefont {Lu}, \citenamefont
  {Goloborodko}, \citenamefont {Abdennur},\ and\ \citenamefont
  {Mirny}}]{Fudenberg2016}%
  \BibitemOpen
  \bibfield  {author} {\bibinfo {author} {\bibfnamefont {G.}~\bibnamefont
  {Fudenberg}}, \bibinfo {author} {\bibfnamefont {M.}~\bibnamefont {Imakaev}},
  \bibinfo {author} {\bibfnamefont {C.}~\bibnamefont {Lu}}, \bibinfo {author}
  {\bibfnamefont {A.}~\bibnamefont {Goloborodko}}, \bibinfo {author}
  {\bibfnamefont {N.}~\bibnamefont {Abdennur}}, \ and\ \bibinfo {author}
  {\bibfnamefont {L.~A.}\ \bibnamefont {Mirny}},\ }\href {\doibase
  10.1101/024620} {\bibfield  {journal} {\bibinfo  {journal} {Cell Rep.}\
  }\textbf {\bibinfo {volume} {15}},\ \bibinfo {pages} {2038} (\bibinfo {year}
  {2016})}\BibitemShut {NoStop}%
\bibitem [{\citenamefont {Wang}\ \emph {et~al.}(2017)\citenamefont {Wang},
  \citenamefont {Brand{\~{a}}o}, \citenamefont {Le}, \citenamefont {Laub},\
  and\ \citenamefont {Rudner}}]{Wang2017c}%
  \BibitemOpen
  \bibfield  {author} {\bibinfo {author} {\bibfnamefont {X.}~\bibnamefont
  {Wang}}, \bibinfo {author} {\bibfnamefont {H.~B.}\ \bibnamefont
  {Brand{\~{a}}o}}, \bibinfo {author} {\bibfnamefont {T.~B.~K.}\ \bibnamefont
  {Le}}, \bibinfo {author} {\bibfnamefont {M.~T.}\ \bibnamefont {Laub}}, \ and\
  \bibinfo {author} {\bibfnamefont {D.~Z.}\ \bibnamefont {Rudner}},\
  }\href@noop {} {\bibfield  {journal} {\bibinfo  {journal} {Science}\ }\textbf
  {\bibinfo {volume} {527}},\ \bibinfo {pages} {524} (\bibinfo {year}
  {2017})}\BibitemShut {NoStop}%
\bibitem [{\citenamefont {Ganji}\ \emph {et~al.}(2018)\citenamefont {Ganji},
  \citenamefont {Shaltiel}, \citenamefont {Bisht}, \citenamefont {Kim},
  \citenamefont {Kalichava}, \citenamefont {Haering},\ and\ \citenamefont
  {Dekker}}]{Ganji2018}%
  \BibitemOpen
  \bibfield  {author} {\bibinfo {author} {\bibfnamefont {M.}~\bibnamefont
  {Ganji}}, \bibinfo {author} {\bibfnamefont {I.~A.}\ \bibnamefont {Shaltiel}},
  \bibinfo {author} {\bibfnamefont {S.}~\bibnamefont {Bisht}}, \bibinfo
  {author} {\bibfnamefont {E.}~\bibnamefont {Kim}}, \bibinfo {author}
  {\bibfnamefont {A.}~\bibnamefont {Kalichava}}, \bibinfo {author}
  {\bibfnamefont {C.~H.}\ \bibnamefont {Haering}}, \ and\ \bibinfo {author}
  {\bibfnamefont {C.}~\bibnamefont {Dekker}},\ }\href {\doibase
  10.1126/science.aar7831} {\bibfield  {journal} {\bibinfo  {journal} {Science
  (80-. ).}\ }\textbf {\bibinfo {volume} {360}},\ \bibinfo {pages} {102}
  (\bibinfo {year} {2018})}\BibitemShut {NoStop}%
\bibitem [{\citenamefont {Davidson}\ \emph {et~al.}(2019)\citenamefont
  {Davidson}, \citenamefont {Bauer}, \citenamefont {Goetz}, \citenamefont
  {Tang}, \citenamefont {Wutz},\ and\ \citenamefont {Peters}}]{Davidson2019a}%
  \BibitemOpen
  \bibfield  {author} {\bibinfo {author} {\bibfnamefont {I.~F.}\ \bibnamefont
  {Davidson}}, \bibinfo {author} {\bibfnamefont {B.}~\bibnamefont {Bauer}},
  \bibinfo {author} {\bibfnamefont {D.}~\bibnamefont {Goetz}}, \bibinfo
  {author} {\bibfnamefont {W.}~\bibnamefont {Tang}}, \bibinfo {author}
  {\bibfnamefont {G.}~\bibnamefont {Wutz}}, \ and\ \bibinfo {author}
  {\bibfnamefont {J.-M.}\ \bibnamefont {Peters}},\ }\href {\doibase
  10.1126/science.aaz3418} {\bibfield  {journal} {\bibinfo  {journal} {Science
  (New York, N.Y.)}\ }\textbf {\bibinfo {volume} {366}},\ \bibinfo {pages}
  {1338} (\bibinfo {year} {2019})}\BibitemShut {NoStop}%
\bibitem [{\citenamefont {Brackley}\ \emph {et~al.}(2017)\citenamefont
  {Brackley}, \citenamefont {Johnson}, \citenamefont {Michieletto},
  \citenamefont {Morozov}, \citenamefont {Nicodemi}, \citenamefont {Cook},\
  and\ \citenamefont {Marenduzzo}}]{Brackley2017prl}%
  \BibitemOpen
  \bibfield  {author} {\bibinfo {author} {\bibfnamefont {C.}~\bibnamefont
  {Brackley}}, \bibinfo {author} {\bibfnamefont {J.}~\bibnamefont {Johnson}},
  \bibinfo {author} {\bibfnamefont {D.}~\bibnamefont {Michieletto}}, \bibinfo
  {author} {\bibfnamefont {A.}~\bibnamefont {Morozov}}, \bibinfo {author}
  {\bibfnamefont {M.}~\bibnamefont {Nicodemi}}, \bibinfo {author}
  {\bibfnamefont {P.}~\bibnamefont {Cook}}, \ and\ \bibinfo {author}
  {\bibfnamefont {D.}~\bibnamefont {Marenduzzo}},\ }\href {\doibase
  10.1103/PhysRevLett.119.138101} {\bibfield  {journal} {\bibinfo  {journal}
  {Phys. Rev. Lett.}\ }\textbf {\bibinfo {volume} {119}},\ \bibinfo {pages}
  {138101} (\bibinfo {year} {2017})}\BibitemShut {NoStop}%
\bibitem [{\citenamefont {Davidson}\ \emph {et~al.}(2016)\citenamefont
  {Davidson}, \citenamefont {Goetz}, \citenamefont {Zaczek}, \citenamefont
  {Molodtsov}, \citenamefont {{Huis in 't Veld}}, \citenamefont {Weissmann},
  \citenamefont {Litos}, \citenamefont {Cisneros}, \citenamefont
  {Ocampo‐Hafalla}, \citenamefont {Ladurner}, \citenamefont {Uhlmann},
  \citenamefont {Vaziri},\ and\ \citenamefont {Peters}}]{Davidson2016}%
  \BibitemOpen
  \bibfield  {author} {\bibinfo {author} {\bibfnamefont {I.~F.}\ \bibnamefont
  {Davidson}}, \bibinfo {author} {\bibfnamefont {D.}~\bibnamefont {Goetz}},
  \bibinfo {author} {\bibfnamefont {M.~P.}\ \bibnamefont {Zaczek}}, \bibinfo
  {author} {\bibfnamefont {M.~I.}\ \bibnamefont {Molodtsov}}, \bibinfo {author}
  {\bibfnamefont {P.~J.}\ \bibnamefont {{Huis in 't Veld}}}, \bibinfo {author}
  {\bibfnamefont {F.}~\bibnamefont {Weissmann}}, \bibinfo {author}
  {\bibfnamefont {G.}~\bibnamefont {Litos}}, \bibinfo {author} {\bibfnamefont
  {D.~A.}\ \bibnamefont {Cisneros}}, \bibinfo {author} {\bibfnamefont
  {M.}~\bibnamefont {Ocampo‐Hafalla}}, \bibinfo {author} {\bibfnamefont
  {R.}~\bibnamefont {Ladurner}}, \bibinfo {author} {\bibfnamefont
  {F.}~\bibnamefont {Uhlmann}}, \bibinfo {author} {\bibfnamefont
  {A.}~\bibnamefont {Vaziri}}, \ and\ \bibinfo {author} {\bibfnamefont
  {J.}~\bibnamefont {Peters}},\ }\href {\doibase 10.15252/embj.201695402}
  {\bibfield  {journal} {\bibinfo  {journal} {EMBO J.}\ }\textbf {\bibinfo
  {volume} {35}},\ \bibinfo {pages} {2671} (\bibinfo {year}
  {2016})}\BibitemShut {NoStop}%
\bibitem [{\citenamefont {Stigler}\ \emph {et~al.}(2016)\citenamefont
  {Stigler}, \citenamefont {{\c{C}}amdere}, \citenamefont {Koshland},\ and\
  \citenamefont {Greene}}]{Stigler2016}%
  \BibitemOpen
  \bibfield  {author} {\bibinfo {author} {\bibfnamefont {J.}~\bibnamefont
  {Stigler}}, \bibinfo {author} {\bibfnamefont {G.~{\"{O}}.}\ \bibnamefont
  {{\c{C}}amdere}}, \bibinfo {author} {\bibfnamefont {D.~E.}\ \bibnamefont
  {Koshland}}, \ and\ \bibinfo {author} {\bibfnamefont {E.~C.}\ \bibnamefont
  {Greene}},\ }\href {\doibase 10.1016/j.celrep.2016.04.003} {\bibfield
  {journal} {\bibinfo  {journal} {Cell Rep.}\ }\textbf {\bibinfo {volume}
  {15}},\ \bibinfo {pages} {988} (\bibinfo {year} {2016})}\BibitemShut
  {NoStop}%
\bibitem [{\citenamefont {Ryu}\ \emph {et~al.}(2020)\citenamefont {Ryu},
  \citenamefont {Bouchoux}, \citenamefont {Liu}, \citenamefont {Kim},
  \citenamefont {Minamino}, \citenamefont {de~Groot}, \citenamefont {Katan},
  \citenamefont {Bonato}, \citenamefont {Marenduzzo}, \citenamefont
  {Michieletto}, \citenamefont {Uhlmann},\ and\ \citenamefont
  {Dekker}}]{Ryu2020}%
  \BibitemOpen
  \bibfield  {author} {\bibinfo {author} {\bibfnamefont {J.-K.}\ \bibnamefont
  {Ryu}}, \bibinfo {author} {\bibfnamefont {C.}~\bibnamefont {Bouchoux}},
  \bibinfo {author} {\bibfnamefont {H.~W.}\ \bibnamefont {Liu}}, \bibinfo
  {author} {\bibfnamefont {E.}~\bibnamefont {Kim}}, \bibinfo {author}
  {\bibfnamefont {M.}~\bibnamefont {Minamino}}, \bibinfo {author}
  {\bibfnamefont {R.}~\bibnamefont {de~Groot}}, \bibinfo {author}
  {\bibfnamefont {A.~J.}\ \bibnamefont {Katan}}, \bibinfo {author}
  {\bibfnamefont {A.}~\bibnamefont {Bonato}}, \bibinfo {author} {\bibfnamefont
  {D.}~\bibnamefont {Marenduzzo}}, \bibinfo {author} {\bibfnamefont
  {D.}~\bibnamefont {Michieletto}}, \bibinfo {author} {\bibfnamefont
  {F.}~\bibnamefont {Uhlmann}}, \ and\ \bibinfo {author} {\bibfnamefont
  {C.}~\bibnamefont {Dekker}},\ }\href {\doibase 10.1101/2020.06.13.149716}
  {\bibfield  {journal} {\bibinfo  {journal} {bioRxiv}\ ,\ \bibinfo {pages}
  {2020.06.13.149716}} (\bibinfo {year} {2020})}\BibitemShut {NoStop}%
\bibitem [{\citenamefont {Nasmyth}(2001)}]{Nasmyth2001}%
  \BibitemOpen
  \bibfield  {author} {\bibinfo {author} {\bibfnamefont {K.}~\bibnamefont
  {Nasmyth}},\ }\href {\doibase 10.1146/annurev.genet.35.102401.091334}
  {\bibfield  {journal} {\bibinfo  {journal} {Annu Rev Gen}\ }\textbf {\bibinfo
  {volume} {35}},\ \bibinfo {pages} {673} (\bibinfo {year} {2001})}\BibitemShut
  {NoStop}%
\bibitem [{\citenamefont {Hirano}\ and\ \citenamefont
  {Hirano}(2002)}]{Hirano2002}%
  \BibitemOpen
  \bibfield  {author} {\bibinfo {author} {\bibfnamefont {M.}~\bibnamefont
  {Hirano}}\ and\ \bibinfo {author} {\bibfnamefont {T.}~\bibnamefont
  {Hirano}},\ }\href {\doibase 10.1093/emboj/cdf575} {\bibfield  {journal}
  {\bibinfo  {journal} {EMBO J.}\ }\textbf {\bibinfo {volume} {21}},\ \bibinfo
  {pages} {5733} (\bibinfo {year} {2002})}\BibitemShut {NoStop}%
\bibitem [{\citenamefont {Gibcus}\ \emph {et~al.}(2018)\citenamefont {Gibcus},
  \citenamefont {Samejima}, \citenamefont {Goloborodko}, \citenamefont
  {Samejima}, \citenamefont {Naumova}, \citenamefont {Nuebler}, \citenamefont
  {Kanemaki}, \citenamefont {Xie}, \citenamefont {Paulson}, \citenamefont
  {Earnshaw}, \citenamefont {Mirny},\ and\ \citenamefont
  {Dekker}}]{Gibcus2018}%
  \BibitemOpen
  \bibfield  {author} {\bibinfo {author} {\bibfnamefont {J.~H.}\ \bibnamefont
  {Gibcus}}, \bibinfo {author} {\bibfnamefont {K.}~\bibnamefont {Samejima}},
  \bibinfo {author} {\bibfnamefont {A.}~\bibnamefont {Goloborodko}}, \bibinfo
  {author} {\bibfnamefont {I.}~\bibnamefont {Samejima}}, \bibinfo {author}
  {\bibfnamefont {N.}~\bibnamefont {Naumova}}, \bibinfo {author} {\bibfnamefont
  {J.}~\bibnamefont {Nuebler}}, \bibinfo {author} {\bibfnamefont {M.~T.}\
  \bibnamefont {Kanemaki}}, \bibinfo {author} {\bibfnamefont {L.}~\bibnamefont
  {Xie}}, \bibinfo {author} {\bibfnamefont {J.~R.}\ \bibnamefont {Paulson}},
  \bibinfo {author} {\bibfnamefont {W.~C.}\ \bibnamefont {Earnshaw}}, \bibinfo
  {author} {\bibfnamefont {L.~A.}\ \bibnamefont {Mirny}}, \ and\ \bibinfo
  {author} {\bibfnamefont {J.}~\bibnamefont {Dekker}},\ }\href {\doibase
  10.1126/science.aao6135} {\bibfield  {journal} {\bibinfo  {journal} {Science
  (80-. ).}\ }\textbf {\bibinfo {volume} {359}} (\bibinfo {year} {2018}),\
  10.1126/science.aao6135}\BibitemShut {NoStop}%
\bibitem [{\citenamefont {Goloborodko}\ \emph {et~al.}(2016)\citenamefont
  {Goloborodko}, \citenamefont {Marko},\ and\ \citenamefont
  {Mirny}}]{Goloborodko2016a}%
  \BibitemOpen
  \bibfield  {author} {\bibinfo {author} {\bibfnamefont {A.}~\bibnamefont
  {Goloborodko}}, \bibinfo {author} {\bibfnamefont {J.~F.}\ \bibnamefont
  {Marko}}, \ and\ \bibinfo {author} {\bibfnamefont {L.~A.}\ \bibnamefont
  {Mirny}},\ }\href {\doibase 10.1016/j.bpj.2016.02.041} {\bibfield  {journal}
  {\bibinfo  {journal} {Biophys. J.}\ }\textbf {\bibinfo {volume} {110}},\
  \bibinfo {pages} {2162} (\bibinfo {year} {2016})}\BibitemShut {NoStop}%
\bibitem [{\citenamefont {Piskadlo}\ and\ \citenamefont
  {Oliveira}(2017)}]{Piskadlo2017}%
  \BibitemOpen
  \bibfield  {author} {\bibinfo {author} {\bibfnamefont {E.}~\bibnamefont
  {Piskadlo}}\ and\ \bibinfo {author} {\bibfnamefont {R.~A.}\ \bibnamefont
  {Oliveira}},\ }\href {\doibase 10.3390/ijms18122751} {\bibfield  {journal}
  {\bibinfo  {journal} {Int. J. Mol. Sci.}\ }\textbf {\bibinfo {volume} {18}},\
  \bibinfo {pages} {1} (\bibinfo {year} {2017})}\BibitemShut {NoStop}%
\bibitem [{\citenamefont {Orlandini}\ \emph {et~al.}(2019)\citenamefont
  {Orlandini}, \citenamefont {Marenduzzo},\ and\ \citenamefont
  {Michieletto}}]{Orlandini2019}%
  \BibitemOpen
  \bibfield  {author} {\bibinfo {author} {\bibfnamefont {E.}~\bibnamefont
  {Orlandini}}, \bibinfo {author} {\bibfnamefont {D.}~\bibnamefont
  {Marenduzzo}}, \ and\ \bibinfo {author} {\bibfnamefont {D.}~\bibnamefont
  {Michieletto}},\ }\href {\doibase 10.1073/pnas.1815394116} {\bibfield
  {journal} {\bibinfo  {journal} {Proc. Natl. Acad. Sci.}\ }\textbf {\bibinfo
  {volume} {116}},\ \bibinfo {pages} {8149} (\bibinfo {year}
  {2019})}\BibitemShut {NoStop}%
\bibitem [{\citenamefont {Inoue}\ \emph {et~al.}(2006)\citenamefont {Inoue},
  \citenamefont {Miyauchi}, \citenamefont {Nakajima}, \citenamefont
  {Takashima}, \citenamefont {Yamaguchi},\ and\ \citenamefont
  {Harada}}]{Inoue2006}%
  \BibitemOpen
  \bibfield  {author} {\bibinfo {author} {\bibfnamefont {Y.}~\bibnamefont
  {Inoue}}, \bibinfo {author} {\bibfnamefont {M.}~\bibnamefont {Miyauchi}},
  \bibinfo {author} {\bibfnamefont {H.}~\bibnamefont {Nakajima}}, \bibinfo
  {author} {\bibfnamefont {Y.}~\bibnamefont {Takashima}}, \bibinfo {author}
  {\bibfnamefont {H.}~\bibnamefont {Yamaguchi}}, \ and\ \bibinfo {author}
  {\bibfnamefont {A.}~\bibnamefont {Harada}},\ }\href {\doibase
  10.1021/ja061095t} {\bibfield  {journal} {\bibinfo  {journal} {J. Am. Chem.
  Soc.}\ }\textbf {\bibinfo {volume} {128}},\ \bibinfo {pages} {8994} (\bibinfo
  {year} {2006})}\BibitemShut {NoStop}%
\bibitem [{\citenamefont {Forgan}\ \emph {et~al.}(2011)\citenamefont {Forgan},
  \citenamefont {Sauvage},\ and\ \citenamefont {Stoddart}}]{Forgan2011}%
  \BibitemOpen
  \bibfield  {author} {\bibinfo {author} {\bibfnamefont {R.~S.}\ \bibnamefont
  {Forgan}}, \bibinfo {author} {\bibfnamefont {J.~P.}\ \bibnamefont {Sauvage}},
  \ and\ \bibinfo {author} {\bibfnamefont {J.~F.}\ \bibnamefont {Stoddart}},\
  }\href {\doibase 10.1021/cr200034u} {\bibfield  {journal} {\bibinfo
  {journal} {Chem. Re}\ }\textbf {\bibinfo {volume} {111}},\ \bibinfo {pages}
  {5434} (\bibinfo {year} {2011})}\BibitemShut {NoStop}%
\bibitem [{\citenamefont {Ito}(2007)}]{Ito2007}%
  \BibitemOpen
  \bibfield  {author} {\bibinfo {author} {\bibfnamefont {K.}~\bibnamefont
  {Ito}},\ }\href {\doibase 10.1295/polymj.PJ2006239} {\bibfield  {journal}
  {\bibinfo  {journal} {Polym. J.}\ }\textbf {\bibinfo {volume} {39}},\
  \bibinfo {pages} {489} (\bibinfo {year} {2007})}\BibitemShut {NoStop}%
\bibitem [{\citenamefont {{Bin Imran}}\ \emph {et~al.}(2014)\citenamefont {{Bin
  Imran}}, \citenamefont {Esaki}, \citenamefont {Gotoh}, \citenamefont {Seki},
  \citenamefont {Ito}, \citenamefont {Sakai},\ and\ \citenamefont
  {Takeoka}}]{BinImran2014}%
  \BibitemOpen
  \bibfield  {author} {\bibinfo {author} {\bibfnamefont {A.}~\bibnamefont {{Bin
  Imran}}}, \bibinfo {author} {\bibfnamefont {K.}~\bibnamefont {Esaki}},
  \bibinfo {author} {\bibfnamefont {H.}~\bibnamefont {Gotoh}}, \bibinfo
  {author} {\bibfnamefont {T.}~\bibnamefont {Seki}}, \bibinfo {author}
  {\bibfnamefont {K.}~\bibnamefont {Ito}}, \bibinfo {author} {\bibfnamefont
  {Y.}~\bibnamefont {Sakai}}, \ and\ \bibinfo {author} {\bibfnamefont
  {Y.}~\bibnamefont {Takeoka}},\ }\href {\doibase 10.1038/ncomms6124}
  {\bibfield  {journal} {\bibinfo  {journal} {Nature Communications}\ }\textbf
  {\bibinfo {volume} {5}},\ \bibinfo {pages} {1} (\bibinfo {year}
  {2014})}\BibitemShut {NoStop}%
\bibitem [{\citenamefont {Yasuda}\ \emph {et~al.}(2019)\citenamefont {Yasuda},
  \citenamefont {Toda}, \citenamefont {Mayumi}, \citenamefont {Yokoyama},
  \citenamefont {Morita},\ and\ \citenamefont {Ito}}]{Yasuda2019}%
  \BibitemOpen
  \bibfield  {author} {\bibinfo {author} {\bibfnamefont {Y.}~\bibnamefont
  {Yasuda}}, \bibinfo {author} {\bibfnamefont {M.}~\bibnamefont {Toda}},
  \bibinfo {author} {\bibfnamefont {K.}~\bibnamefont {Mayumi}}, \bibinfo
  {author} {\bibfnamefont {H.}~\bibnamefont {Yokoyama}}, \bibinfo {author}
  {\bibfnamefont {H.}~\bibnamefont {Morita}}, \ and\ \bibinfo {author}
  {\bibfnamefont {K.}~\bibnamefont {Ito}},\ }\href {\doibase
  10.1021/acs.macromol.9b00118} {\bibfield  {journal} {\bibinfo  {journal}
  {Macromolecules}\ }\textbf {\bibinfo {volume} {52}},\ \bibinfo {pages} {3787}
  (\bibinfo {year} {2019})}\BibitemShut {NoStop}%
\bibitem [{\citenamefont {Zandi}\ \emph {et~al.}(2003)\citenamefont {Zandi},
  \citenamefont {Kantor},\ and\ \citenamefont {Kardar}}]{Zandi2003}%
  \BibitemOpen
  \bibfield  {author} {\bibinfo {author} {\bibfnamefont {R.}~\bibnamefont
  {Zandi}}, \bibinfo {author} {\bibfnamefont {Y.}~\bibnamefont {Kantor}}, \
  and\ \bibinfo {author} {\bibfnamefont {M.}~\bibnamefont {Kardar}},\ }\href
  {http://arxiv.org/abs/cond-mat/0306587} {\ \textbf {\bibinfo {volume} {1}},\
  \bibinfo {pages} {1} (\bibinfo {year} {2003})}\BibitemShut {NoStop}%
\bibitem [{\citenamefont {Duplantier}(1989)}]{Duplantier1989}%
  \BibitemOpen
  \bibfield  {author} {\bibinfo {author} {\bibfnamefont {B.}~\bibnamefont
  {Duplantier}},\ }\href {\doibase 10.1007/BF01019770} {\bibfield  {journal}
  {\bibinfo  {journal} {J. Stat. Phys.}\ }\textbf {\bibinfo {volume} {54}},\
  \bibinfo {pages} {581} (\bibinfo {year} {1989})}\BibitemShut {NoStop}%
\bibitem [{not()}]{noteknot}%
  \BibitemOpen
  \href@noop {} {}\bibinfo {note} {This is because the scaling of the number of
  configurations of a knotted loop is known only in the asymptotic limit of a
  tight knot on a large loop.}\BibitemShut {Stop}%
\bibitem [{\citenamefont {Bonato}\ \emph {et~al.}(2020)\citenamefont {Bonato},
  \citenamefont {Brackley}, \citenamefont {Johnson}, \citenamefont
  {Michieletto},\ and\ \citenamefont {Mareduzzo}}]{Bonato2020}%
  \BibitemOpen
  \bibfield  {author} {\bibinfo {author} {\bibfnamefont {A.}~\bibnamefont
  {Bonato}}, \bibinfo {author} {\bibfnamefont {C.~A.}\ \bibnamefont
  {Brackley}}, \bibinfo {author} {\bibfnamefont {J.}~\bibnamefont {Johnson}},
  \bibinfo {author} {\bibfnamefont {D.}~\bibnamefont {Michieletto}}, \ and\
  \bibinfo {author} {\bibfnamefont {D.}~\bibnamefont {Mareduzzo}},\ }\href
  {\doibase 10.1039/c9sm01875a} {\bibfield  {journal} {\bibinfo  {journal}
  {Soft Matter}\ ,\ \bibinfo {pages} {2406}} (\bibinfo {year}
  {2020})}\BibitemShut {NoStop}%
\bibitem [{\citenamefont {Vian}\ \emph {et~al.}(2018)\citenamefont {Vian},
  \citenamefont {Pekowska}, \citenamefont {Rao}, \citenamefont {Levens},
  \citenamefont {{Lieberman Aiden}}, \citenamefont {Casellas}, \citenamefont
  {Vian}, \citenamefont {Kieffer-Kwon}, \citenamefont {Jung}, \citenamefont
  {Baranello}, \citenamefont {Huang}, \citenamefont {{El Khattabi}},
  \citenamefont {Dose}, \citenamefont {Pruett}, \citenamefont {Sanborn},
  \citenamefont {Canela}, \citenamefont {Maman}, \citenamefont {Oksanen},
  \citenamefont {Resch}, \citenamefont {Li}, \citenamefont {Lee}, \citenamefont
  {Kovalchuk}, \citenamefont {Tang}, \citenamefont {Nelson}, \citenamefont {{Di
  Pierro}}, \citenamefont {Cheng}, \citenamefont {Machol}, \citenamefont
  {{Glenn St Hilaire}}, \citenamefont {Durand}, \citenamefont {Shamim},
  \citenamefont {Stamenova}, \citenamefont {Onuchic}, \citenamefont {Ruan},\
  and\ \citenamefont {Nussenzweig}}]{Vian2018a}%
  \BibitemOpen
  \bibfield  {author} {\bibinfo {author} {\bibfnamefont {L.}~\bibnamefont
  {Vian}}, \bibinfo {author} {\bibfnamefont {A.}~\bibnamefont {Pekowska}},
  \bibinfo {author} {\bibfnamefont {S.~S.}\ \bibnamefont {Rao}}, \bibinfo
  {author} {\bibfnamefont {D.}~\bibnamefont {Levens}}, \bibinfo {author}
  {\bibfnamefont {E.}~\bibnamefont {{Lieberman Aiden}}}, \bibinfo {author}
  {\bibfnamefont {R.}~\bibnamefont {Casellas}}, \bibinfo {author}
  {\bibfnamefont {L.}~\bibnamefont {Vian}}, \bibinfo {author} {\bibfnamefont
  {K.-R.}\ \bibnamefont {Kieffer-Kwon}}, \bibinfo {author} {\bibfnamefont
  {S.}~\bibnamefont {Jung}}, \bibinfo {author} {\bibfnamefont {L.}~\bibnamefont
  {Baranello}}, \bibinfo {author} {\bibfnamefont {S.-C.}\ \bibnamefont
  {Huang}}, \bibinfo {author} {\bibfnamefont {L.}~\bibnamefont {{El
  Khattabi}}}, \bibinfo {author} {\bibfnamefont {M.}~\bibnamefont {Dose}},
  \bibinfo {author} {\bibfnamefont {N.}~\bibnamefont {Pruett}}, \bibinfo
  {author} {\bibfnamefont {A.~L.}\ \bibnamefont {Sanborn}}, \bibinfo {author}
  {\bibfnamefont {A.}~\bibnamefont {Canela}}, \bibinfo {author} {\bibfnamefont
  {Y.}~\bibnamefont {Maman}}, \bibinfo {author} {\bibfnamefont
  {A.}~\bibnamefont {Oksanen}}, \bibinfo {author} {\bibfnamefont
  {W.}~\bibnamefont {Resch}}, \bibinfo {author} {\bibfnamefont
  {X.}~\bibnamefont {Li}}, \bibinfo {author} {\bibfnamefont {B.}~\bibnamefont
  {Lee}}, \bibinfo {author} {\bibfnamefont {A.~L.}\ \bibnamefont {Kovalchuk}},
  \bibinfo {author} {\bibfnamefont {Z.}~\bibnamefont {Tang}}, \bibinfo {author}
  {\bibfnamefont {S.}~\bibnamefont {Nelson}}, \bibinfo {author} {\bibfnamefont
  {M.}~\bibnamefont {{Di Pierro}}}, \bibinfo {author} {\bibfnamefont {R.~R.}\
  \bibnamefont {Cheng}}, \bibinfo {author} {\bibfnamefont {I.}~\bibnamefont
  {Machol}}, \bibinfo {author} {\bibfnamefont {B.}~\bibnamefont {{Glenn St
  Hilaire}}}, \bibinfo {author} {\bibfnamefont {N.~C.}\ \bibnamefont {Durand}},
  \bibinfo {author} {\bibfnamefont {M.~S.}\ \bibnamefont {Shamim}}, \bibinfo
  {author} {\bibfnamefont {E.~K.}\ \bibnamefont {Stamenova}}, \bibinfo {author}
  {\bibfnamefont {J.~N.}\ \bibnamefont {Onuchic}}, \bibinfo {author}
  {\bibfnamefont {Y.}~\bibnamefont {Ruan}}, \ and\ \bibinfo {author}
  {\bibfnamefont {A.}~\bibnamefont {Nussenzweig}},\ }\href {\doibase
  10.1016/j.cell.2018.03.072} {\bibfield  {journal} {\bibinfo  {journal}
  {Cell}\ }\textbf {\bibinfo {volume} {173}},\ \bibinfo {pages} {1} (\bibinfo
  {year} {2018})}\BibitemShut {NoStop}%
\bibitem [{\citenamefont {Uusk{\"{u}}la-Reimand}\ \emph
  {et~al.}(2016)\citenamefont {Uusk{\"{u}}la-Reimand}, \citenamefont {Hou},
  \citenamefont {Samavarchi-Tehrani}, \citenamefont {Rudan}, \citenamefont
  {Liang}, \citenamefont {Medina-Rivera}, \citenamefont {Mohammed},
  \citenamefont {Schmidt}, \citenamefont {Schwalie}, \citenamefont {Young},
  \citenamefont {Reimand}, \citenamefont {Hadjur}, \citenamefont {Gingras},\
  and\ \citenamefont {Wilson}}]{Uuskula-Reimand2016}%
  \BibitemOpen
  \bibfield  {author} {\bibinfo {author} {\bibfnamefont {L.}~\bibnamefont
  {Uusk{\"{u}}la-Reimand}}, \bibinfo {author} {\bibfnamefont {H.}~\bibnamefont
  {Hou}}, \bibinfo {author} {\bibfnamefont {P.}~\bibnamefont
  {Samavarchi-Tehrani}}, \bibinfo {author} {\bibfnamefont {M.~V.}\ \bibnamefont
  {Rudan}}, \bibinfo {author} {\bibfnamefont {M.}~\bibnamefont {Liang}},
  \bibinfo {author} {\bibfnamefont {A.}~\bibnamefont {Medina-Rivera}}, \bibinfo
  {author} {\bibfnamefont {H.}~\bibnamefont {Mohammed}}, \bibinfo {author}
  {\bibfnamefont {D.}~\bibnamefont {Schmidt}}, \bibinfo {author} {\bibfnamefont
  {P.}~\bibnamefont {Schwalie}}, \bibinfo {author} {\bibfnamefont {E.~J.}\
  \bibnamefont {Young}}, \bibinfo {author} {\bibfnamefont {J.}~\bibnamefont
  {Reimand}}, \bibinfo {author} {\bibfnamefont {S.}~\bibnamefont {Hadjur}},
  \bibinfo {author} {\bibfnamefont {A.-C.}\ \bibnamefont {Gingras}}, \ and\
  \bibinfo {author} {\bibfnamefont {M.~D.}\ \bibnamefont {Wilson}},\ }\href
  {\doibase 10.1186/s13059-016-1043-8} {\bibfield  {journal} {\bibinfo
  {journal} {Genome Biol}\ }\textbf {\bibinfo {volume} {17}},\ \bibinfo {pages}
  {1} (\bibinfo {year} {2016})}\BibitemShut {NoStop}%
\bibitem [{\citenamefont {Pippenger}(1989)}]{Pippenger1989}%
  \BibitemOpen
  \bibfield  {author} {\bibinfo {author} {\bibfnamefont {N.}~\bibnamefont
  {Pippenger}},\ }\href
  {http://www.sciencedirect.com/science/article/pii/0166218X8990005X}
  {\bibfield  {journal} {\bibinfo  {journal} {Discret. Appl. Math.}\ }\textbf
  {\bibinfo {volume} {25}},\ \bibinfo {pages} {273} (\bibinfo {year}
  {1989})}\BibitemShut {NoStop}%
\bibitem [{\citenamefont {Micheletti}\ \emph {et~al.}(2006)\citenamefont
  {Micheletti}, \citenamefont {Marenduzzo}, \citenamefont {Orlandini},\ and\
  \citenamefont {Summers}}]{Micheletti2006c}%
  \BibitemOpen
  \bibfield  {author} {\bibinfo {author} {\bibfnamefont {C.}~\bibnamefont
  {Micheletti}}, \bibinfo {author} {\bibfnamefont {D.}~\bibnamefont
  {Marenduzzo}}, \bibinfo {author} {\bibfnamefont {E.}~\bibnamefont
  {Orlandini}}, \ and\ \bibinfo {author} {\bibfnamefont {D.~W.}\ \bibnamefont
  {Summers}},\ }\href {\doibase 10.1063/1.2162886} {\bibfield  {journal}
  {\bibinfo  {journal} {Journal of Chemical Physics}\ }\textbf {\bibinfo
  {volume} {124}} (\bibinfo {year} {2006}),\ 10.1063/1.2162886}\BibitemShut
  {NoStop}%
\bibitem [{\citenamefont {Nakahata}\ \emph {et~al.}(2016)\citenamefont
  {Nakahata}, \citenamefont {Mori}, \citenamefont {Takashima}, \citenamefont
  {Yamaguchi},\ and\ \citenamefont {Harada}}]{Nakahata2016}%
  \BibitemOpen
  \bibfield  {author} {\bibinfo {author} {\bibfnamefont {M.}~\bibnamefont
  {Nakahata}}, \bibinfo {author} {\bibfnamefont {S.}~\bibnamefont {Mori}},
  \bibinfo {author} {\bibfnamefont {Y.}~\bibnamefont {Takashima}}, \bibinfo
  {author} {\bibfnamefont {H.}~\bibnamefont {Yamaguchi}}, \ and\ \bibinfo
  {author} {\bibfnamefont {A.}~\bibnamefont {Harada}},\ }\href {\doibase
  10.1016/j.chempr.2016.09.013} {\bibfield  {journal} {\bibinfo  {journal}
  {Chem}\ }\textbf {\bibinfo {volume} {1}},\ \bibinfo {pages} {766} (\bibinfo
  {year} {2016})}\BibitemShut {NoStop}%
\end{thebibliography}%


\begin{thebibliography}{4}%
\makeatletter
\providecommand \@ifxundefined [1]{%
 \@ifx{#1\undefined}
}%
\providecommand \@ifnum [1]{%
 \ifnum #1\expandafter \@firstoftwo
 \else \expandafter \@secondoftwo
 \fi
}%
\providecommand \@ifx [1]{%
 \ifx #1\expandafter \@firstoftwo
 \else \expandafter \@secondoftwo
 \fi
}%
\providecommand \natexlab [1]{#1}%
\providecommand \enquote  [1]{``#1''}%
\providecommand \bibnamefont  [1]{#1}%
\providecommand \bibfnamefont [1]{#1}%
\providecommand \citenamefont [1]{#1}%
\providecommand \href@noop [0]{\@secondoftwo}%
\providecommand \href [0]{\begingroup \@sanitize@url \@href}%
\providecommand \@href[1]{\@@startlink{#1}\@@href}%
\providecommand \@@href[1]{\endgroup#1\@@endlink}%
\providecommand \@sanitize@url [0]{\catcode `\\12\catcode `\$12\catcode
  `\&12\catcode `\#12\catcode `\^12\catcode `\_12\catcode `\%12\relax}%
\providecommand \@@startlink[1]{}%
\providecommand \@@endlink[0]{}%
\providecommand \url  [0]{\begingroup\@sanitize@url \@url }%
\providecommand \@url [1]{\endgroup\@href {#1}{\urlprefix }}%
\providecommand \urlprefix  [0]{URL }%
\providecommand \Eprint [0]{\href }%
\providecommand \doibase [0]{http://dx.doi.org/}%
\providecommand \selectlanguage [0]{\@gobble}%
\providecommand \bibinfo  [0]{\@secondoftwo}%
\providecommand \bibfield  [0]{\@secondoftwo}%
\providecommand \translation [1]{[#1]}%
\providecommand \BibitemOpen [0]{}%
\providecommand \bibitemStop [0]{}%
\providecommand \bibitemNoStop [0]{.\EOS\space}%
\providecommand \EOS [0]{\spacefactor3000\relax}%
\providecommand \BibitemShut  [1]{\csname bibitem#1\endcsname}%
\let\auto@bib@innerbib\@empty
\bibitem [{\citenamefont {Zandi}\ \emph {et~al.}(2003)\citenamefont {Zandi},
  \citenamefont {Kantor},\ and\ \citenamefont {Kardar}}]{Zandi2003}%
  \BibitemOpen
  \bibfield  {author} {\bibinfo {author} {\bibfnamefont {R.}~\bibnamefont
  {Zandi}}, \bibinfo {author} {\bibfnamefont {Y.}~\bibnamefont {Kantor}}, \
  and\ \bibinfo {author} {\bibfnamefont {M.}~\bibnamefont {Kardar}},\ }\href
  {http://arxiv.org/abs/cond-mat/0306587} {\ \textbf {\bibinfo {volume} {1}},\
  \bibinfo {pages} {1} (\bibinfo {year} {2003})}\BibitemShut {NoStop}%
\bibitem [{\citenamefont {Bonato}\ \emph {et~al.}(2020)\citenamefont {Bonato},
  \citenamefont {Brackley}, \citenamefont {Johnson}, \citenamefont
  {Michieletto},\ and\ \citenamefont {Mareduzzo}}]{Bonato2020}%
  \BibitemOpen
  \bibfield  {author} {\bibinfo {author} {\bibfnamefont {A.}~\bibnamefont
  {Bonato}}, \bibinfo {author} {\bibfnamefont {C.~A.}\ \bibnamefont
  {Brackley}}, \bibinfo {author} {\bibfnamefont {J.}~\bibnamefont {Johnson}},
  \bibinfo {author} {\bibfnamefont {D.}~\bibnamefont {Michieletto}}, \ and\
  \bibinfo {author} {\bibfnamefont {D.}~\bibnamefont {Mareduzzo}},\ }\href
  {\doibase 10.1039/c9sm01875a} {\bibfield  {journal} {\bibinfo  {journal}
  {Soft Matter}\ ,\ \bibinfo {pages} {2406}} (\bibinfo {year}
  {2020})}\BibitemShut {NoStop}%
\bibitem [{\citenamefont {Duplantier}(1989)}]{Duplantier1989}%
  \BibitemOpen
  \bibfield  {author} {\bibinfo {author} {\bibfnamefont {B.}~\bibnamefont
  {Duplantier}},\ }\href {\doibase 10.1007/BF01019770} {\bibfield  {journal}
  {\bibinfo  {journal} {J. Stat. Phys.}\ }\textbf {\bibinfo {volume} {54}},\
  \bibinfo {pages} {581} (\bibinfo {year} {1989})}\BibitemShut {NoStop}%
\bibitem [{\citenamefont {Metzler}\ \emph {et~al.}(2002)\citenamefont
  {Metzler}, \citenamefont {Hanke}, \citenamefont {Dommersnes}, \citenamefont
  {Kantor},\ and\ \citenamefont {Kardar}}]{Metzler2002}%
  \BibitemOpen
  \bibfield  {author} {\bibinfo {author} {\bibfnamefont {R.}~\bibnamefont
  {Metzler}}, \bibinfo {author} {\bibfnamefont {A.}~\bibnamefont {Hanke}},
  \bibinfo {author} {\bibfnamefont {P.~G.}\ \bibnamefont {Dommersnes}},
  \bibinfo {author} {\bibfnamefont {Y.}~\bibnamefont {Kantor}}, \ and\ \bibinfo
  {author} {\bibfnamefont {M.}~\bibnamefont {Kardar}},\ }\href {\doibase
  10.1103/PhysRevE.65.061103} {\bibfield  {journal} {\bibinfo  {journal} {Phys.
  Rev. E}\ }\textbf {\bibinfo {volume} {65}},\ \bibinfo {pages} {1} (\bibinfo
  {year} {2002})}\BibitemShut {NoStop}%
\end{thebibliography}%

\end{document}


\title{Simplifying Topological Entanglements by Entropic Competition of Slip-Links: Supplemental Material}

\author{Andrea Bonato}
\address{University of Edinburgh, SUPA, School of Physics and Astronomy, Peter Guthrie Road, EH9 3FD, Edinburgh, UK }

\author{Davide Marenduzzo}
\address{University of Edinburgh, SUPA, School of Physics and Astronomy, Peter Guthrie Road, EH9 3FD, Edinburgh, UK }

\author{Davide Michieletto}
\affiliation{University of Edinburgh, SUPA, School of Physics and Astronomy, Peter Guthrie Road, EH9 3FD, Edinburgh, UK }
\affiliation{MRC Human Genetics Unit, Institute of Genetics and Molecular Medicine, University of Edinburgh, Edinburgh EH4 2XU, UK}

\maketitle

\renewcommand{\theequation}{S\arabic{equation}}

\renewcommand{\thefigure}{S\arabic{figure}}

\section*{Model and methods}

We perform Molecular Dynamics simulations of bead-and-spring polymers made up of beads of size $\sigma$. The simulation box is separated into two equally large portions by an impenetrable wall and each bead interacting with the wall is subject to a repulsive force perpendicular to the wall described by the potential:
\begin{equation}\label{eq:wallpot}
U_{\rm wall}(r) = \left\{
\begin{array}{lr}
\epsilon \left[ \frac{2}{15}\left(\frac{\sigma}{r}\right)^{9} - \left(\frac{\sigma}{r}\right)^3 -\frac{2}{3}\sqrt{\frac{5}{2}} \right] & \, r \le r_c \\
0 & \, r > r_c
\end{array} \right. \, ,
\end{equation}
where $r$ is the separation between the bead and the wall and $r_c=\left(\frac{2}{5}\right)^{\frac16}\sigma$.
Beads of the polymer are strung together by finite-extension-nonlinear-elastic (FENE) bonds:
\begin{equation}\label{eq:Ufene}
U_{\rm FENE}(r) = \left\{
\begin{array}{lcl}
-0.5kR_0^2 \ln\left(1-(r / R_0)^2\right) & \ r\le R_0 \\ \infty & \
r> R_0 &
\end{array} \right. \, ,
\end{equation}
where $R_{0}=1.5\sigma$ is the maximum extension of the bond; interactions between beads are governed by the Weeks-Chandler-Andersen (WCA) potential:
\begin{equation}\label{eq:WCA}
U_{\rm WCA}(r) = \left\{
\begin{array}{lr}
4 \epsilon \left[ \left(\frac{\sigma}{r}\right)^{12} - \left(\frac{\sigma}{r}\right)^6 + \frac14 \right] & \, r \le r_c \\
0 & \, r > r_c
\end{array} \right. \, ,
\end{equation}
where $r$ is the distance between the bead centres and $r_c=2^{1/6}\sigma$. 
To account for the effect of stiffness (arising, for instance, if the polymers represent chromatin fibres), we introduced a Kratky-Porod potential acting on triplets of neighbouring beads:
\begin{equation}
U_{\rm bend} = \dfrac{k_BT l_p}{\sigma}(1 + \cos{\theta})
\label{eq:KPpot},
\end{equation}
where $\theta$ is the angle between consecutive bonds. The initial configurations are designed such that a polymer passes through two holes in a wall, the size of which is small enough to let just one monomer through at a time. In this way, the polymer is divided into two segments, physically separated by the wall. Additionally, we fix the position of one of the two beads closest to the wall so that the segments can exchange monomers only through the other hole. This set up is reminiscent of the one used to study entropic competition in Ref.~\cite{Zandi2003}. Slip-links are modelled as physical square hand-cuffs of side $4\sigma$, held together by four FENE bounds and opened in a ``planar" arrangement (this is achieved with the potential described by Eq.~\eqref{eq:KPpot} and large enough persistence length).
To study the kinetics of topological simplification, we allow a short ($10$ beads) segment to undergo strand-crossings. This is achieved by setting to $0$ the amplitude of the (WCA) potential regulating the interaction between the beads of this segment and each bead composing the rest of the polymer, excluding the two immediately next to the segment boundaries. In our slip-link vs knot simulations, this segment is kept in the knot side, by placing it next to the fixed bead (wall).

\section{Note on the Correlation Time}
It is important that with our simulations we can sample the systems ergodically and avoid kinetic traps which may exist in MD simulations. To make sure of this the simulations are sampled every $10^6$ LAMMPS time steps, comparable to the de-correlation time of a polymer of length up to at least $200$ monomers. Within this time also the SLs loaded onto the substrate have travelled a significant amount of contour length~\cite{Bonato2020} and so we assume that each sampled configuration is independent of others. Since they were obtained by simulating polymers of length $500$, the histograms shown in the second panels of lines B C and D of figure~\ref{fig:SL_vs_SL}, suffer from longer de-correlation times. Improving the statistics would, however, require much longer simulations.

\section{Equations for SL vs SL}

\subsection{Duplantier Scaling for Polymer Networks}

The number of configurations of an asymptotically long monodisperse -- i.e., where branches are equal in length -- polymer network of total length $L$ and topology $\mathcal{G}$ in the dilute regime scales as~\cite{Duplantier1989}
\begin{equation}
\mathcal{Z}_\mathcal{G}(L) \sim \mu^L L^{\gamma_{\mathcal{G}}-1},
\label{eq:scaling_dilute}
\end{equation}
where
\begin{equation}
\gamma_{\mathcal{G}}-1=-\nu d\mathcal{L} + \sum_{M\geq 1}n_M\sigma_M
\label{eq:gamma_dilute}
\end{equation}
is a topology dependent exponent: $d$ is the dimension, $\mathcal{L}$ is the number of loops, $n_M$ the number of vertices of multeplicity -- i.e., the number of branches connected to the vertex -- $M$ and ${\sigma_M}$ the star polymers exponents.

If the network is grafted to a surface, Eq.~\eqref{eq:scaling_dilute} changes to~\cite{Duplantier1989}
\begin{equation}
\mathcal{Z}^S_\mathcal{G}(L) \sim \mu^L L^{\gamma^S_{\mathcal{G}}-1},
\label{eq:scaling_surface}
\end{equation}
where
\begin{equation}
\gamma^S_{\mathcal{G}}-1=-\nu (d\mathcal{L} +\mathcal{V_S}-1+\mathcal{L}_S)+ \sum_{M\geq 1}(n_M\sigma_M+n^S_M\sigma^S_M).
\label{eq:gamma_surface}
\end{equation}
$n^S_M$ is the number of surface vertices (those grafted to the surface) of multeplicity $M$, $\mathcal{V_S}=\sum_{M\geq 1}n^S_M$ is the number of surface vertices, $\mathcal{L}_S=\sum_{M\geq 1}Mn^S_M$ is the number of branches connected to the surface and ${\sigma^S_M}$ are the surface star polymers exponents.

\subsection{Surface loop and surface round table exponents}
Let us call surface loop ($\mathcal{O}$) a simple loop with a single vertex grafted to a surface. A surface loop and the ``surface'' round table (RT) and necklace (with 2 slip-links only, NC) topologies we considered in our simulations (see Fig~\ref{fig:SL_vs_SL}) have only one surface vertex of multeplicity 2: $\mathcal{V}_S=1$ and $\mathcal{L}_S=2$. According to Duplantier~\cite{Duplantier1989},
\begin{equation}
\sigma^S_2=2\nu-1,
\end{equation}
hence
\begin{equation}
\gamma^S_{\mathcal{O}}-1=-d\nu -1
\label{eq:gamma_sloop}
\end{equation}
and
\begin{equation}
\gamma^S_{RT,N_{SL}}-1=\gamma_{RT,N_{SL}}-2=-d\nu\mathcal(N_{SL}+1)+\sigma_4 N_{SL}-1,
\label{eq:gamma_RT}
\end{equation}
where $N_{SL}$ is the number of slip-links.
\begin{equation}
\gamma^S_{NC,2}=\gamma^S_{RT,2}.
\label{eq:gamma_NC}
\end{equation}

\subsection{Surface round table and necklace scaling with the lengths of the loops}
Analogously to what predicted for the RT networks without surface~\cite{Metzler2002}, we assume the partition function of the surface RT networks with $N_{SL}$ slip-links scales as
\begin{equation}
\begin{split}
\mathcal{Z}_{RT,N_{SL}}(L)\sim&\mu^LA(s_1,\dots,s_{N_{SL}},L)\left(L-\sum_{i=1}^{N_{SL}}s_i\right)^{\gamma^S_{RT,N_{SL}}-1} \\
&\times \mathcal{X}\left(\frac{s_1}{L-\sum_{i=1}^{N_{SL}}s_i},\dots,\frac{s_{N_{SL}}}{L-\sum_{i=1}^{N_{SL}}s_i}\right)
\end{split}
\label{eq:singlel_scaling_2}
\end{equation} 
where with $s_i$ we indicate the lengths of the loops subtended by the slip-links (slip-linked loops) and $A(s_1,\dots,s_{N_{SL}},L)$ accounts for the sliding entropy of the slip-links.
\newline
\paragraph{Round table}
We now suppose the peripheral slip-linked loops are much smaller than the internal loop, as we see in our simulations. They contribute equally to Eq.~\eqref{eq:singlel_scaling_2}:
\begin{equation}
\begin{split}
\mathcal{Z}_{RT,N_{SL}}(L)\sim\mu^LA(s_1,\dots,s_{N_{SL}},&L)\left(L-\sum_{i=1}^{N_{SL}}s_i\right)^{\gamma^S_{RT,N_{SL}}-1}\\
&\times\prod_{i=1}^{N_{SL}}\left(\frac{s_i}{L-\sum_{j=1}^{N_{SL}}s_j}\right)^\alpha
\label{eq:singlel_scaling_3}
\end{split}
\end{equation} 
Furthermore, on the basis of this assumption, we require the scaling contribution of the largest loop to be that of a surface loop (without considering the sliding entropy; see~\cite{Metzler2002} for the case without surface). This sets $\alpha$ to
\begin{equation}
\alpha = \frac{\gamma^S_{RT,N_{SL}}-1}{N_{SL}} -\frac{-d\nu-1}{N_{SL}} = -d\nu+\sigma_4,
\label{eq:scaling_perif_loop}
\end{equation}
which is the same loop closure factor for the RT topologies without surface. Finally,
\begin{equation}
\begin{split}
\mathcal{Z}_{RT,N_{SL}}(L)\sim\mu^LA(s_1,\dots,s_{N_{SL}},&L)\left(L-\sum_{i=1}^{N_{SL}}s_i\right)^{-d\nu-1}\\
&\times\prod_{i=1}^{N_{SL}}s_i^{-d\nu+\sigma_4}.
\end{split}
\label{eq:singlel_RT1}
\end{equation}
\ \newline
\paragraph{Surface necklace with $2$ SLs}
In the case of the necklace topology with $2$ slip-links, we write
\begin{equation}
\begin{split}
\mathcal{Z}_{NC,2}(L)\sim&\mu^LB(s_1,s_2,L) (L-s_1-s_2)^{\gamma^S_{NC,2}-1}\\
&\times\left(\frac{s_1}{L-s_1-s_2}\right)^{\alpha_1}\left(\frac{s_2}{L-s_1-s_2}\right)^{\alpha_2},
\end{split}
\label{eq:singlel_scaling_4}
\end{equation} 
where we labelled as $1$ the slip-linked loop grafted to the wall and where $B(s_1,s_2,L)$ accounts for the sliding entropy. This time, we can not state $\alpha_1=\alpha_2$ a priori. However, if we require the scaling contribution of the internal loop (which we assume to be the largest) to be that of a surface loop, and the contribution of the sliding slip-linked loop to be the usual closure factor (Eq.~\eqref{eq:scaling_perif_loop}), the final result is
\begin{equation}
\small
\mathcal{Z}_{NC,2}(L)\sim\mu^LB(s_1,s_2,L)\left(L-s_1-s_2\right)^{-d\nu-1}s_1^{-d\nu+\sigma_4}s_2^{-d\nu+\sigma_4}.
\label{eq:singlel_RT2}
\end{equation}
\newline
Notice the consistency of Eqs.~\ref{eq:singlel_RT1}~and~\ref{eq:singlel_RT2} with the hypothesis of the peripheral loops being small depends on the form of the sliding entropy terms $A$ and $B$. For the surface necklace this might be a point of discussion; nonetheless, Eq.~\eqref{eq:singlel_RT2} represents our simulated data reasonably well (see Fig.~2B in the main text).

\section{Single SL vs single SL}

\begin{figure}[t!]
	\includegraphics[width=0.45\textwidth]{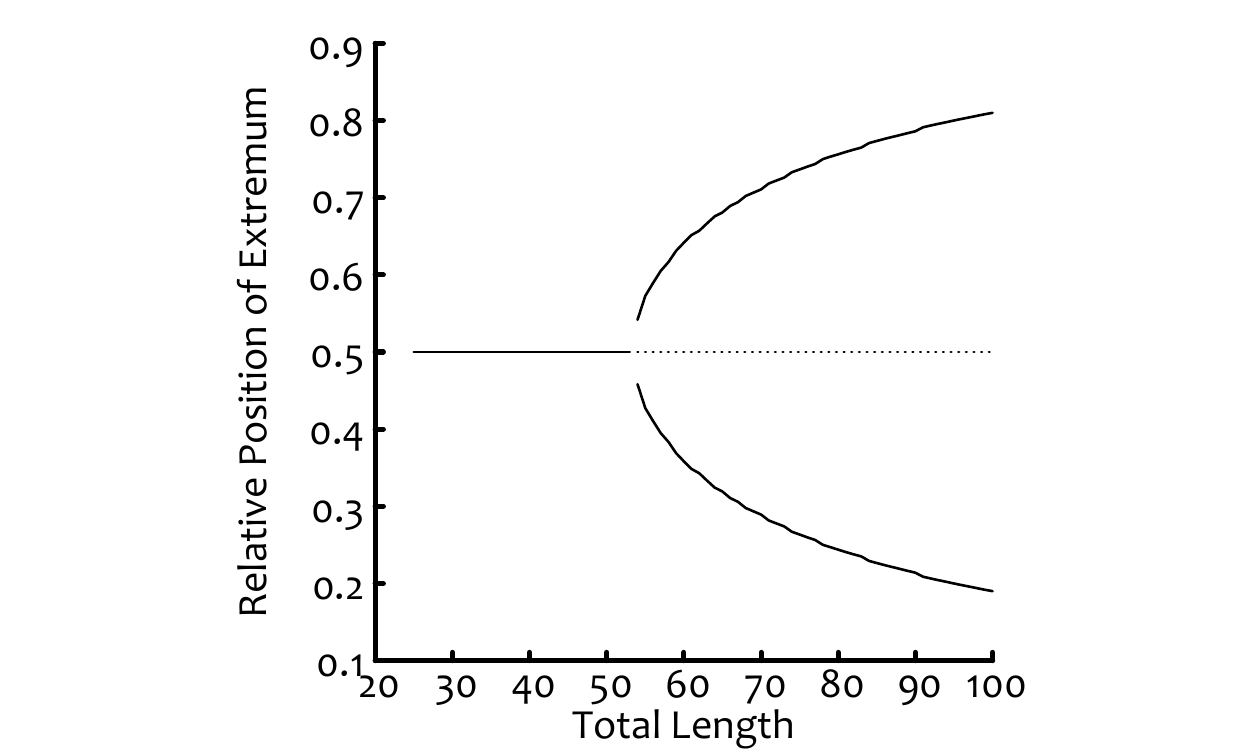}
	\caption{Transition from unimodal to bimodal shape in the distribution of lengths of two slip-links competing with each other as predicted by Eq.~\eqref{eq:scaling1SLvs1SL}. The black solid lines are the positions of the stable points in the the length distribution of one side of a polymer loop loaded with two competing slip-links. Dashed lines represent the unstable points.}
	\label{fig:SL_vs_SL_SI}
\end{figure}

The number of configurations of two slip-links competing with each other (Fig.~\ref{fig:SL_vs_SL}A) scales as:
	\begin{equation}
	\begin{split}
	\mu^L&\int_{s^{m}_1}^{l-s^{m}_2-s^{m}_3}{ds_1}\left(l-s^{m}_2-s^{m}_3\right) \frac{s_1^{-3\nu+\sigma_4}s_2^{-3\nu+\sigma_4}}{\left(l-s_1 \right)^{3\nu+1}} \;\times \; \\ &\int_{s^{m}_1}^{L-l-s^{m}_2-s^{m}_3}{ds_1}\left(L-l-s^{m}_2-s^{m}_3\right) \frac{s_1^{-3\nu+\sigma_4}s_2^{-3\nu+\sigma_4}}{\left(L-l-s_1 \right)^{3\nu+1}}, \\
	\end{split}
	\label{eq:scaling1SLvs1SL}
\end{equation}
where the coefficients $s^{m}_i$ account for the physical size of the slip-links: they are the minimum size of the segments of the slip-linked polymer; $L$ and $l$ are, respectively, the total length and the length of one side of the polymer. Notice the integrals account for the sliding entropy of the slip-links.\\
Figure~\ref{fig:SL_vs_SL_SI} is obtained numerically from Eq.~\eqref{eq:scaling1SLvs1SL}. 

\section{$2 $SLs vs $2$ SLs}
The number of configurations of two consecutive slip-links competing with two consecutive slip-links (Fig.~\ref{fig:SL_vs_SL}C) scales as:
\begin{widetext}
	\begin{equation*}
	\begin{split}
	\mu^L&\int_{s^{m}_1}^{l-s^{m}_2-s^{m}_3-s^{m}_4-s^{m}_5}{ds_1}\int_{s^{m}_2}^{l-s_1-s^{m}_3-s^{m}_4-s^{m}_5}{ds_2} \left(l-s_1-s_2-s^{m}_3-s^{m}_4-s^{m}_5\right)^2 \frac{s_1^{-3\nu+\sigma_4}s_2^{-3\nu+\sigma_4}}{\left(l-s_1-s_2 \right)^{3\nu+1}} \;\times \; \\ &\int_{s^{m}_1}^{L-l-s^{m}_2-s^{m}_3-s^{m}_4-s^{m}_5}{ds_1}\int_{s^{m}_2}^{L-l-s_1-s^{m}_3-s^{m}_4-s^{m}_5}{ds_2} \left(L-l-s_1-s_2-s^{m}_3-s^{m}_4-s^{m}_5\right)^2 \frac{s_1^{-3\nu+\sigma_4}s_2^{-3\nu+\sigma_4}}{\left(L-l-s_1-s_2 \right)^{3\nu+1}}.
	\end{split}
	\label{eq:scaling2SLvs2SLSeries}
	\end{equation*}
\end{widetext}

\subsubsection*{$2$ nested SLs vs $2$ nested SLs}
The number of configurations of two nested slip-links competing with two nested slip-links (Fig.~\ref{fig:SL_vs_SL}B) scales as:

\begin{widetext}
	\begin{equation}
\begin{split}
\mu^L&\left[\int_{s^{m}_1}^{l-s^{m}_2-s^{m}_3-s^{m}_4-s^{m}_5}{ds_1}\int_{s^{m}_2}^{l-s_1-s^{m}_3-s^{m}_4-s^{m}_5}{ds_2}\int_{s^{m}_3}^{l-s_1-s_2-s^{m}_4-s^{m}_5}{ds_3}\left(l-s_1-s_2-s_3-s^{m}_4-s^{m}_5\right)\right.\\
&\left. \frac{\left(s_1+s_2\right)^{-3\nu+\sigma_4}s_3^{-3\nu+\sigma_4}}{\left(l-s_1-s_2-s_3 \right)^{3\nu+1}} \right] \;\times \; \left[ \int_{s^{m}_1}^{L-l-s^{m}_2-s^{m}_3-s^{m}_4-s^{m}_5}{ds_1}\int_{s^{m}_2}^{L-l-s_1-s^{m}_3-s^{m}_4-s^{m}_5}{ds_2}\int_{s^{m}_3}^{L-l-s_1-s_2-s^{m}_4-s^{m}_5}{ds_3} \right.\\
&\left(L-l-s_1-s_2-s_3-s^{m}_4-s^{m}_5\right)\left. \frac{\left(s_1+s_2\right)^{-3\nu+\sigma_4}s_3^{-3\nu+\sigma_4}}{\left(L-l-s_1-s_2-s_3 \right)^{3\nu+1}}\right].
\end{split}
\label{eq:scaling2SLvs2SLNested}
\end{equation}
\end{widetext}

\begin{figure*}[t!]
	\includegraphics[width=0.95\textwidth]{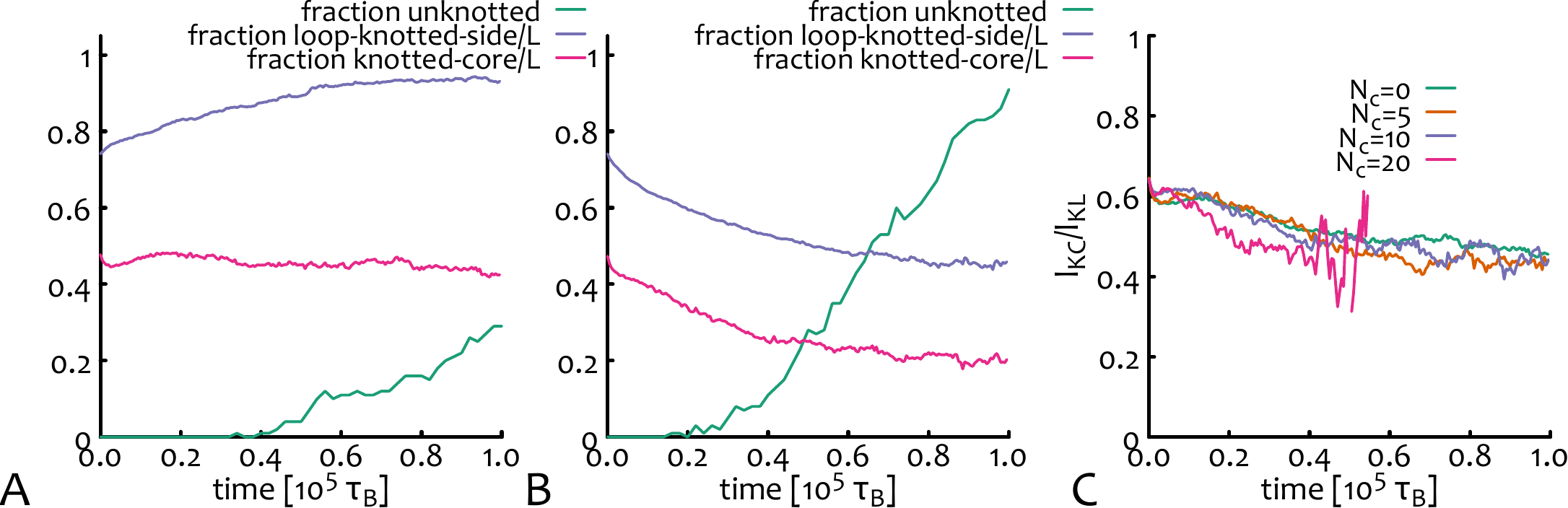}
	\caption{\textbf{The knotted core follows the knotted loop.} \textbf{A-B} Show the time evolution of the average loop length on the knotted side (blue), of the average length of knot core (pink, the minimum length that can be found knotted) and the fraction of simulations that have simplified the knot (green) for \textbf{A} $N_{SL}=0$ and \textbf{B} $N_{SL}=10$. One can notice that the case with $N_{SL}=0$ shows a weak swelling of the knotted side (blue curve increases), but the knot core remains quite constant (pink curve is flat) and the simulations simplify the topology slowly over time (green curve increases slowly). On the other hand,   $N_{SL}=10$ shows a much faster topological simplification rate (green curve sharply increases) and the length of knotted side (and core) are concomitantly reduced. \textbf{C} The ratio of the knotted core length over the knotted loop length remains around $1/2$ for the whole simulation and for different choices of $N_{SL}$, indicating that as the knotted loop looses contour length, the knot core is compressed. }
	\label{fig:unknotting_SI}
\end{figure*}

\bibliographystyle{apsrev4-1}
\bibliography{library}